\documentstyle[epsfig]{mn}
\input{epsf}

\newcommand{\pl}{\partial}
\renewcommand{\d}{{\rm d}}
\newcommand{\beq}{\begin{equation}}
\newcommand{\eeq}{\end{equation}}
\newcommand{\beqa}{\begin{eqnarray}} 
\newcommand{\eeqa}{\end{eqnarray}}
\newcommand{\bea}{\begin{array}} 
\newcommand{\ea}{\end{array}} 
\newcommand{\lag}{\langle}
\newcommand{\rag}{\rangle}
\newcommand{\Om}{\Omega_{\rm m}}
\newcommand{\Ol}{\Omega_{\Lambda}}
\newcommand{\De}{{\cal D}}
\newcommand{\gam}{\gamma}
\newcommand{\cP}{{\cal P}}
\newcommand{\kappamin}{\kappa_{\rm min}}
\newcommand{\kappah}{\hat{\kappa}}
\newcommand{\wh}{\hat{w}}
\newcommand{\kaphs}{\hat{\kappa}_s}
\newcommand{\Map}{M_{\rm ap}}
\newcommand{\tS}{\tilde{S}}
\newcommand{\bx}{{\bf x}}
\newcommand{\bk}{{\bf k}}
\newcommand{\kpar}{k_{\parallel}}
\newcommand{\kperp}{\bk_{\perp}}
\newcommand{\kperpDt}{k_{\perp}\De\theta_s}

\newcommand{\inta}{\int_{-i\infty}^{+i\infty}}
\newcommand{\rhob}{\overline{\rho}}
\newcommand{\xib}{\overline{\xi}}
\newcommand{\phikap}{\varphi_{\kaphs}}

\newcommand{\dum}{s}
\newcommand{\xidum}{\xi_{\dum}}
\newcommand{\phidum}{\varphi_{\dum}}
\newcommand{\Ikap}{I_{\kappa}}
\newcommand{\xikap}{\xi_{\kaphs}}
\newcommand{\om}{\omega}
\newcommand{\omb}{\overline{\omega}}
\newcommand{\Istel}{I_{\kappa *}}
\newcommand{\Skewkaps}{S_3^{\kappa_s}}
\newcommand{\Skewkaphs}{S_3^{\kaphs}}

\title[Weak lensing convergence]
{From linear to non-linear scales: analytical and numerical predictions for the
weak lensing convergence}
\author[Barber A. J. et al.]
{Andrew J. Barber$^{1}$,  
Dipak Munshi$^{2,3}$, Patrick Valageas$^{4}$\\
$^{1}$Astronomy Centre, University of Sussex, Falmer, Brighton, BN1 9QJ,
United Kingdom\\
$^{2}$Institute of Astronomy, Madingley Road,
Cambridge, CB3 OHA, United Kingdom\\
$^{3}$Astrophysics Group, Cavendish Laboratory, Madingley Road, 
Cambridge CB3 OHE, United Kingdom\\
$^{4}$ Service de Physique Th\'eorique, 
CEA Saclay, 91191 Gif-sur-Yvette, France \\
}

\begin{document}
\maketitle

\begin{abstract}
Weak lensing convergence can be used directly to map and probe the
dark mass distribution in the universe. Building on earlier studies,
we recall how the statistics of the convergence field are related to
the statistics of the underlying mass distribution, in particular to
the many-body density correlations. We describe two model-independent
approximations which provide two simple methods to compute the
probability distribution function, pdf, of the convergence. We apply
one of these to the case where the density field can be described by a
log-normal pdf.  Next, we discuss two hierarchical models for the
high-order correlations which allow one to perform exact calculations
and evaluate the previous approximations in such specific
cases. Finally, we apply these methods to a very simple model for the
evolution of the density field from linear to highly non-linear
scales. Comparisons with the results obtained from numerical
simulations, obtained from a number of different realizations, show
excellent agreement with our theoretical predictions. We have probed
various angular scales in the numerical work and considered sources at
14 different redshifts in each of two different cosmological
scenarios, an open cosmology and a flat cosmology with non-zero
cosmological constant. Our simulation technique employs computations
of the full 3-d shear matrices along the line of sight from the source
redshift to the observer and is complementary to more popular
ray-tracing algorithms. Our results therefore provide a valuable
cross-check for such complementary simulation techniques, as well as
for our simple analytical model, from the linear to the highly
non-linear regime.
\end{abstract}

 

\section{Introduction}

Weak gravitational lensing is responsible for the shearing and
magnification in the images of high-redshift sources due to the
presence of intervening mass. Since the lensing effects arise from
deflections of the light rays due to fluctuations of the gravitational
potential, they can be directly related to the underlying density
field of the large-scale structure and not necessarily to the presence
of luminous structure in the form of galaxies and
clusters. Consequently, statistical analysis of observed weak lensing
data has proved to be a powerful tool to probe the underlying density,
which is assumed to be dominated by the dark matter content. Since the
underlying density field and its evolution is strongly dependent on
the cosmological parameters, observational surveys have been
particularly fruitful in providing estimates of the parameters, which
can be pre-set in analytical calculations or in numerical
simulations. 


Analytical computations for weak lensing statistics can be readily
obtained for large smoothing angles, $> 10'$, where perturbative
calculations are applicable (e.g., Villumsen, 1996, Stebbins, 1996,
Bernardeau et al., 1997, Jain \& Seljak, 1997, Kaiser, 1998, Van
Waerbeke, Bernardeau \& Mellier, 1999, and Schneider et al.,
1998). However, on small angular scales, $< 5'$, especially relevant
to observational surveys with small sky coverage, perturbative
calculations are no longer valid and models to represent the
gravitational clustering in the non-linear regime have had to be
devised.

Thus, a number of useful techniques have been developed to transform
from the linear matter power spectrum to a fully non-linear
spectrum. Originally, Hamilton et al. (1991) considered the evolution
of the matter correlation function and this work was extended by
Peacock \& Dodds (1996) to describe the non-linear evolution of the
matter power spectrum. Their methods were based on the conservation of
mass and a rescaling of physical lengths in the different regimes,
following the ``stable-clustering'' {\it Ansatz} which assumes that
small scales decouple and are statistically frozen in proper
coordinates (Peebles, 1980).

In a different approach, Peacock \& Smith (2000) and Seljak (2000)
developed a model based on the random distribution of dark matter
haloes, modulated by the large-scale matter distribution. This ``Halo
Model'' for non-linear evolution is able to reproduce the matter power
spectrum of $N$-body simulations over a wide range of scales and has
the advantage of relating the linear and non-linear power at the same
scale through fitting formul\ae. Smith et al. (2002) have presented a
new set of fitting functions based on the Halo Model and calibrated to
a set of $N$-body simulations. Barber \& Taylor (2002) have shown
excellent agreement between the power spectrum in the lensing
convergence obtained from numerical simulations and the predictions
from the Halo model fitting functions of Smith et al. (2002).

An advantage of the halo model is that in principle it can predict the
higher-order correlation functions in the highly non-linear regime. Indeed,
on small scales the latter are set by the density profile of the halos
as the $p-$point correlation is dominated by the contribution associated
with all $p$ points being within the same halo. However, the neglect of
substructures may lead to larger inaccuracies for higher-order statistics.
Moreover, at intermediate scales one also probes the correlations among
different halos which introduces new unknowns and makes explicit calculations
cumbersome. Finally, such a model for the density field is not 
well-suited to describe low-density and underdense regions (e.g., voids,
filaments) which are outside virialized objects.

These problems have motivated the recourse to an alternative approach
which is directly based on the many-body correlations. The most common
model of this kind is to express the $p-$point correlations as a sum
of $p-1$ products over the two-point correlations linking all $p$
points. This yields the class of ``Hierarchical models,'' which are
specified by the weights given to any such topology associated with
the $p-1$ products (e.g., Fry 1984, Schaeffer 1984, Bernardeau \&
Schaeffer 1992, and Szapudi \& Szalay 1993, 1997). Once these weights
have been assigned it is possible to resum all many-body correlations
and to compute the probability distribution function (pdf) of the
density field, or of any quantity which is linearly dependent on the
matter density. This is most easily done for ``minimal tree-models'',
where the weight associated to a given tree-topology is set by its
vertices (e.g., Bernardeau \& Schaeffer 1992), or for ``stellar
models'', which only contain stellar diagrams (Valageas, Barber \&
Munshi, 2003). Such an approach is very well-suited to the study of
weak-lensing which only involves the matter density field and does not
make the distinction between astrophysical objects like clusters,
Lyman-$\alpha$ clouds or voids. Hence the many-body correlations of
the density field are precisely the quantities of interest which
directly appear through the statistics of weak-lensing effects, rather
than the possible decomposition of the density field over different
classes of objects.

Using such an approach, coupled to the Hamilton et al. (1991) prescription 
for the two-point correlation, Valageas (2000a, b) and 
Munshi \& Jain (2000 and 2001) were able to compute the pdf of 
the weak-lensing convergence
whilst the associated bias was considered by Munshi (2000) and the
cumulant correlators associated with such distributions were evaluated
by Munshi \& Jain (2000). Munshi \& Wang (2003) further extended these
studies to cosmological scenarios including dark energy. These methods can
also handle more intricate quantities like the aperture-mass or the shear
which involve compensated filters and require a detailed model for the 
many-body correlations. Thus, using a minimal tree-model for the 
non-linear regime, Bernardeau and Valageas (2000) were able to predict 
the pdf of the aperture-mass and to obtain a good agreement with numerical 
simulations (they also showed that similar techniques could be applied to
the quasi-linear regime where the calculations can actually be made 
rigorous). On the other hand, adopting a stellar model for the many-body 
correlations, Valageas, Barber \& Munshi (2003) obtained excellent agreement 
for the shear pdf when compared with the results of $N$-body simulations. 


A clear alternative to the analytical approaches for the understanding
of weak lensing has been provided by the development of numerical
techniques to simulate different cosmological scenarios. A recent
initiative, designed specifically for weak lensing studies in
numerical simulations, was developed by Couchman, Barber \& Thomas
(1999) whose technique allows the computation of the full
3-dimensional shear matrices at locations along the lines of sight
entirely within each simulation volume. These intermediate matrices
are then combined using the full implementation of the multiple
lens-plane theory (described fully by Schneider, Ehlers \& Falco,
1992) to obtain the final Jacobian matrix for each line of sight,
appropriate for sources at the selected redshift. Excellent agreement
between the results obtained using Couchman et al.'s (1999) method and
analytical predictions have been reported. Barber (2002) has compared
the redshift and scale dependence of the shear variance with the
analytical program of Van Waerbeke et al. (2001), based on fitting
formul\ae. Barber \& Taylor (2002) computed the angular power spectra
for the convergence and magnification and obtained higher-order
moments for the convergence; their results were in close agreement
with the predictions from a halo model, using the functional fitting
devised by Smith et al. (2002). Most recently, Valageas, Barber \&
Munshi (2003) were able to confirm the predictions of their
hierarchical {\em Ansatz}, using a stellar model, for the full pdf of
the shear for sources at redshift 1 and on angular scales
representative of the non-linear regime.


The main goals of the current paper are to test against numerical simulations
a very simple model for the
evolution of the density field which allows one to compute the statistics of
the weak-lensing convergence from linear to highly non-linear scales and to  
compare the various approximations one can use within this framework.
To achieve this we present results for a
selection of source redshifts and for angular scales from $1'.0$ to
$8'.0$, taking us from the quasi-linear regime into the highly non-linear 
regime. We concentrate here on the full pdf and low-order moments for the 
lensing convergence. We present the following analytical approximations.

\begin{enumerate}

\item  We approximate the mean of the many-body correlations over the
   cone, represented by the angular window, by the mean evaluated over
   spherical shells. This approximation is as described by Valageas
   (2000b) and it is completely model-independent. Indeed, it can be applied
   to any description used for the density field. 

\item  Adding a further approximation to the previous method, we can 
   approximate the generating function, $\phikap(y)$, for the 
   smoothed normalised convergence, $\kaphs$, (from which the pdf is
   obtained after performing a Laplace transform) by the
   generating function, $\varphi(y)$, associated with the
   density contrast, $\delta_R$, at the typical scale and redshift probed
   by observations. As shown in Valageas (2000b), this
   procedure is sufficient to recover the properties of the
   convergence with a reasonable accuracy and it is again model-independent.

\item  We apply the previous approximation to the specific case where the
   pdf of the density contrast can be described by a log-normal law.

\item  We briefly recall how to compute exactly the pdf of the convergence
   in the case where the connected correlations of the density field are
   given by a minimal-tree model, following Valageas (2000b) or 
   Bernardeau \& Valageas (2000).

\item  We show how to compute the statistics of the convergence from the 
   stellar model for the connected correlations of the
   density field introduced in Valageas, Barber \& Munshi (2003) to derive
   the pdf of the weak-lensing shear.

Eventually we apply these methods to a very simple model for the
evolution of the density field from the linear to the highly
non-linear regime.

\end{enumerate}

Finally, the analytical predictions are compared with the results of
numerical simulations, in which Couchman et al.'s (1999) fully
3-dimensional shear method is used. The numerical data have been
obtained in two different cosmologies which we will describe as LCDM
(a flat cosmology with a cosmological constant) and OCDM (an open
cosmology with zero cosmological constant). The data have been
computed for sources at 14 different redshifts and on a wide range of
angular scales.

This paper is organized as follows. Section 2 defines our notations
and outlines the basic equations which express the weak-lensing convergence
in terms of the density field along the line of sight.
Section 3 provides a short description of our analytical model for the
evolution of the density field from linear to highly non-linear scales.
We also recall the general relationship between the pdf of the density 
contrast and its cumulant generating function and we briefly discuss
two hierarchical models which can be used to describe the many-body 
density correlations.
Section 4 provides the main analytical results and highlights
various approximations which can be used to derive the statistics
of convergence maps. 
Section 5 describes our simulations and section 6
contains a detailed comparison of simulation results with analytical 
predictions. 
Finally section 7 is reserved for discussion of our results
within the context of observational programme and future plans.

\section{The convergence field}
\label{The convergence field}

As is well-known, the image of a distant source at a redshift $z_s$
received by an observer at redshift $z=0$ is distorted by the
deflection of light due to density fluctuations along the line of
sight. This shearing and magnification are described in terms of
angular position vector, ${\vec \vartheta}$, by the shear tensor $\pl
\delta{\vec \vartheta}/\pl {\vec \vartheta}$, whose trace, $\kappa$,
(also called the convergence) yields the magnification of the source
(in the weak-lensing regime). The convergence is given by (Bernardeau
et al., 1997, and Kaiser, 1998):
\beq
\kappa \simeq \frac{3\Om}{2} \int_0^{\chi_s} \d\chi \; w(\chi,\chi_s) 
\; \delta(\chi).
\label{kappa}
\eeq
In this equation, $\Om$ is the matter density parameter, $\chi$ and
$\chi_s$ are the radial distance along the line of sight and radial 
distance to the source,
respectively, $\delta(\bx)=(\rho(\bx)-\rhob)/\rhob$ is the density
contrast at position $\bx$ and
\beq
w(\chi,\chi_s) = \frac{H_0^2}{c^2} \; \frac{\De(\chi) \De(\chi_s-\chi)}
{\De(\chi_s)} \; (1+z) 
\label{w}
\eeq
ensures the integration takes account of the relevant angular diameter
distances $\De(\chi)$ to radial distance $\chi$, $\De(\chi_s)$ to the
radial position of the source and $\De(\chi_s - \chi)$ from $\chi$ 
to $\chi_s$; 
\beq
\d\chi = \frac{\frac{c}{H_0} \; \d z}{\sqrt{\Ol+(1-\Om-\Ol)(1+z)^2+
\Om(1+z)^3}} ,
\label{chi}
\eeq
where $c$ is the velocity of light, $H_0$ is the present-day value of
the Hubble parameter and $\Ol$ is the vacuum energy density parameter;
finally,
\beq
\De(\chi) = \frac{ \frac{c}{H_0} \sin_K \left( \mid 1-\Om-\Ol \mid^{1/2} 
H_0 \; \chi/c \right) } {\mid 1-\Om-\Ol \mid^{1/2}},
\label{De}
\eeq
in which $\sin_K$ means the hyperbolic sine, sinh, if $(1-\Om-\Ol) > 0$,
or sine if $(1-\Om-\Ol) < 0$; if $(1-\Om-\Ol) = 0$, then $\De(\chi) = \chi$.
Eq.(\ref{kappa}) assumes that the components of the shear tensor are small so
that we can use the Born approximation (i.e. the integral over redshift is 
taken along the unperturbed line of sight) but the density fluctuations 
$\delta$ can be large (Kaiser, 1992). Next, we can see from eq.(\ref{kappa}) 
that there is a minimum value, $\kappamin(z_s)$, for the convergence of 
a source located at redshift $z_s$, which corresponds to an ``empty'' beam 
between the source and the observer ($\delta=-1$ everywhere along the line 
of sight):
\beq
\kappamin = - \frac{3\Om}{2} \int_0^{\chi_s} \d\chi \; w(\chi,\chi_s) .
\label{kappamin}
\eeq
Following Valageas (2000a, b) and Munshi \& Jain (2000) it is convenient to define the ``normalised'' 
convergence, $\kappah$, by:
\beq
\kappah = \frac{\kappa}{|\kappamin|} = \int_0^{\chi_s} \d\chi \; \wh \; 
\delta , \hspace{0.2cm} \mbox{with} \hspace{0.2cm} 
\wh=\frac{w(\chi,\chi_s)}{\int_0^{\chi_s} \d\chi \; w(\chi,\chi_s)} ,
\label{kappah}
\eeq
which obeys $\kappah \geq -1$. Here we introduced the ``normalised selection 
function,'' $\wh(\chi,\chi_s)$. As shown in Valageas (2000a, b), one interest 
of working with normalised quantities like $\kappah$ is that most of the 
cosmological dependence (on $\Om,\Ol$ and $z_s$) and the projection effects 
are encapsulated within $\kappamin$, while the statistics of $\kappah$ 
(e.g., its pdf) mainly probe the deviations from Gaussianity of the density 
field which arise from the non-linear dynamics of gravitational clustering
as well as the amplitude of the density fluctuations ($\sigma_8$). 
If one smoothes the observations with a top-hat window in real space of 
small angular radius, $\theta_s$, one rather considers the filtered 
normalised convergence $\kaphs$ (where the subscript ``s'' refers to 
``smoothed''):
\beq
\kaphs = \int_0^{\theta_s} \frac{\d {\vec \vartheta}}{\pi \theta_s^2} 
\int_0^{\chi_s} \d\chi \; \wh(\chi,\chi_s) \; 
\delta \left( \chi, \De {\vec \vartheta} \right) .
\label{kap1}
\eeq
Here ${\vec \vartheta}$ is a vector in the plane perpendicular to the line 
of sight (we restrict ourselves to small angular windows) over which we 
integrate within the disk $|{\vec \vartheta}| \leq \theta_s$;
we note this by 
the short notation $\int_0^{\theta_s}$. Thus $\chi$ is the radial coordinate,
while $\De {\vec \vartheta}$ is the two-dimensional vector of transverse 
coordinates. Eq.(\ref{kap1}) clearly shows that the smoothed convergence 
$\kaphs$ is actually an average of the density contrast over the cone of 
angular radius $\theta_s$. 

For some purposes it is convenient to work in Fourier space. Thus, we 
define the Fourier transform of the density contrast by:
\beq
\delta({\bx}) = \int \d\bk \; e^{i \bk.\bx} \; \delta(\bk)
\label{deltak}
\eeq
where $\bx$ and $\bk$ are comoving coordinates. Then, eq.(\ref{kap1}) also 
reads:
\beq
\kaphs = \int_0^{\chi_s} \d\chi \; \wh(\chi,\chi_s) \int \d\bk \; 
e^{i \kpar \chi} \; W(\kperpDt) \; \delta( {\bf k} ) ,
\label{kapk1}
\eeq
where $\kpar$ is the component of $\bk$ parallel to the line of sight
and $\kperp$ is the two-dimensional vector formed by the components of 
$\bk$ perpendicular to the line of sight. Here we introduced the Fourier 
form $W(\kperpDt)$ of the real-space top-hat filter of angular radius 
$\theta_s$:
\beq
W(\kperpDt) = \int_0^{\theta_s} \frac{\d {\vec \vartheta}}{\pi \theta_s^2} 
\; e^{i \kperp . \De {\vec \vartheta}} = \frac{2 J_1(\kperpDt)}{\kperpDt} ,
\label{Wk}
\eeq
where $J_1$ is the Bessel function of the first kind of order 1. If we 
choose another filter (e.g., a Gaussian window rather than a top-hat) the 
expression (\ref{kapk1}) remains valid and we simply need to use the relevant 
Fourier window $W(\kperpDt)$. In Fourier space, we also define the 
power-spectrum, $P(k)$, of the density contrast by:
\beq
\lag \delta(\bk_1) \delta(\bk_2) \rag = \delta_D(\bk_1+\bk_2) \; P(k_1) ,
\label{Pk}
\eeq
where $\delta_D$ is Dirac's distribution. Then, we obtain:
\beq
\xi_2(x) = \lag \delta(\bx_1)\delta(\bx_1+\bx) \rag = 
\int \d\bk \; e^{i\bk.\bx} \; P(k) ,
\label{xiPk}
\eeq
for the two-point correlation $\xi_2(x)$ of the density contrast.

\section{Analytical description of the density field}
\label{Analytical models for the density field}

In order to derive the pdf of the convergence, $\cP(\kappa)$,
we clearly need to specify the properties of the underlying density field.
As seen from eq.(\ref{kappa}), the convergence, $\kappa$, is given
by a linear integral along the line of sight over the density field.
Then, the direct computation of the pdf of such a sum over
the redshift of the lenses would yield an infinite number of
convolution products which makes it intractable. However, this problem
can be greatly simplified by working with the logarithm of the Laplace
transform of the pdf, $\varphi(y)$. Indeed, since the Laplace transform 
changes convolutions into ordinary products and the logarithm changes products
into sums, the generating functions $\varphi(y)$ simply add when different 
layers along the line of sight are superposed. This is the basis of the 
method introduced in Valageas (2000a, b). This approach has already been 
presented in detail in various works, for the convergence (e.g.,
Valageas, 2000a, b; Munshi \& Jain, 2000 and 2001), 
the aperture mass (Bernardeau \& Valageas, 2000) and the shear 
(Valageas, Barber \& Munshi, 2003), using various models for the
density field which are well-suited to such a technique. 

Therefore, we 
briefly review in the following sub-sections the various phenomenological 
models which have been put forward to describe the density field, which we
aim to compare in this article through their implications for the pdf of 
the convergence, $\cP(\kappa)$.

\subsection{Cumulant generating function for the density contrast}
\label{Cumulant generating function for the density contrast}

As recalled above, in order to handle the projection effects associated
with the integration of the density fluctuations along the line of sight,
it is convenient to work with the logarithm $\varphi(y)$ of the Laplace 
transform of the pdf. Therefore, it is useful to define also the pdf 
$\cP(\delta_R)$ of the density contrast at scale $R$ through its generating 
function $\varphi(y)$:
\beq
e^{-\varphi(y)/\xib_2} = \int_{-1}^{\infty} \d\delta_R \; 
e^{-\delta_R y/\xib_2} \; \cP(\delta_R) ,
\label{phidel1}
\eeq
where $\delta_R$ is the density contrast within spherical cells of radius $R$
and volume $V$ while $\xib_2$ is its variance:
\beq
\delta_R = \int_V \frac{\d\bx}{V} \; \delta(\bx) \hspace{0.3cm} \mbox{and}
\hspace{0.3cm} \xib_2 = \lag \delta_R^2 \rag .
\label{deltaR1}
\eeq
The pdf $\cP(\delta_R)$ can be recovered from $\varphi(y)$ through the 
inverse Laplace transform:
\beq
\cP(\delta_R) = \inta \frac{\d y}{2\pi i \xib_2} \; 
e^{[\delta_R y - \varphi(y)] /\xib_2} .
\label{Pdel1}
\eeq
As is well-known, the function $\varphi(y)$ defined from eq.(\ref{phidel1})
is also the generating function of the cumulants of the density contrast
(see any textbook on probability theory). Thus, the expansion of $\varphi(y)$
at $y=0$ reads:
\beq
\varphi(y) = \sum_{p=2}^{\infty} \frac{(-1)^{p-1}}{p!} \; S_p \; y^p 
\hspace{0.3cm} \mbox{with} \hspace{0.3cm} 
S_p = \frac{\lag\delta_R^p\rag_c}{\xib_2^{\; p-1}} .
\label{phidel2}
\eeq
The reason it is useful to introduce the variance $\xib_2$ in the 
definition (\ref{phidel1}) of the cumulant generating function $\varphi(y)$
is that it removes most of the dependence on scale and time of the properties
of the density field. More precisely, it can be shown that $\varphi(y)$ 
as defined above has a finite limit in the limit $\xib_2 \rightarrow 0$, which
corresponds to the quasi-linear regime. This exact result can be obtained 
from the expansion (\ref{phidel2}) through a perturbative method 
(Bernardeau, 1994) or more rigorously from eq.(\ref{phidel1}) through
a steepest-descent method (Valageas, 2002). In particular, the derivation
of the generating function $\varphi(y)$ in this quasi-linear limit yields
the implicit system:
\beqa
\varphi(y) & = & y \left[ \zeta(\tau)-\frac{\tau\;\zeta'(\tau)}{2} \right] 
\label{phiMF} \\
\tau & = & - y \; \zeta'(\tau)
\label{tauMF}
\eeqa
where the function $\zeta(\tau)$ is closely related to the spherical dynamics
for the non-linear density contrast (Bernardeau, 1994; Valageas, 2002).
This function $\zeta(\tau)$ only depends on the local slope $n$ of the 
power-spectrum of the density fluctuations and on the cosmological parameters
$\Om(z),\Ol(z)$. However, the dependence on $\Om,\Ol$, is rather 
small (Bernardeau, 1992) so that over the whole quasi-linear regime 
($\xib_2 \la 1$) the pdf $\cP(\delta_R)$ can be fully described through two
quantities only: the variance $\xib_2$ and the local slope $n$ (which yields
$\zeta(\tau)$, whence $\varphi(y)$ and finally $\cP(\delta_R)$ using 
$\xib_2$).

In the non-linear regime there are no more rigorous results for the behaviour
of the pdf $\cP(\delta_R)$. However, a reasonable approximation is provided
by the ``stable-clustering {\it Ansatz}'' (e.g., Peebles 1980; 
Balian \& Schaeffer 1989) where the cumulants obey the scaling law
$\lag\delta_R^p\rag_c \propto \xib_2^{\; p-1}$. Therefore, the coefficients
$S_p$ introduced in eq.(\ref{phidel2}) are again constant with time so that the
generating function $\varphi(y)$ is again independent of $\xib_2$ (i.e. it is
unique for the whole highly non-linear regime $\xib_2 \gg 1$). However, it
should still depend on the local slope $n$ of the power-spectrum. Note that
deviations from the stable-clustering {\it Ansatz} simply mean that 
$\varphi(y)$ should still exhibit a weak dependence on time within the 
highly non-linear regime.

\subsection{A simple parameterization for the generating function $\varphi(y)$}
\label{A simple parameterization for the generating function}

From the properties of the generating function $\varphi(y)$ recalled above
in Sect.~\ref{Cumulant generating function for the density contrast}, we
choose the following simple parameterization for $\varphi(y)$. For all regimes,
from quasi-linear scales down to highly non-linear scales, we parameterize
$\varphi(y)$ through the associated function $\zeta(\tau)$ which obeys the 
implicit system (\ref{phiMF})-(\ref{tauMF}). Next, we choose a simple 
phenomenological prescription for $\zeta(\tau)$. Following previous works 
(e.g., Bernardeau \& Schaeffer 1992, Bernardeau \& Valageas 2000, 
Valageas, Barber \& Munshi, 2003) we use:
\beq
\zeta(\tau)= \left (1+\frac{\tau}{\kappa}\right)^{-\kappa} - 1 ,
\label{zetadef}
\eeq
where we have kept the usual notation, $\kappa$ (not to be confused with the 
convergence), for the free parameter which enters the definition of 
$\zeta(\tau)$. In order to handle the variation of $\varphi(y)$ (hence of
$\zeta(\tau)$) from linear to highly non-linear scales, we let this parameter
$\kappa$ vary so as to recover the correct skewness $S_3$ of the density
contrast. From eqs.(\ref{phiMF})-(\ref{tauMF}) this yields:
\beq
\kappa= \frac{3}{S_3-3} .
\label{S3kappa}
\eeq
For $S_3>3$ we have $\kappa>0$ while for $0<S_3<3$ we have $\kappa<-1$.
In the limit $S_3 \rightarrow 3$ the function $\zeta(\tau)$ defined in
eq.(\ref{zetadef}) goes to $\zeta(\tau) \rightarrow e^{-\tau}-1$ (and
$|\kappa| \rightarrow \infty$). Then, we take for the skewness 
in the highly non-linear regime, $S_3^{\rm NL}$, the prediction of HEPT 
(Scoccimarro \& Frieman, 1999) and for the quasi-linear regime, 
$S_3^{\rm QL}$, the exact result obtained from perturbative theory:
\beq
S_3^{\rm NL} = 3 \frac{4-2^n}{1+2^{n+1}} , \hspace{0.3cm} 
S_3^{\rm QL} = \frac{34}{7} - (n+3) .
\label{S31}
\eeq
Here $n$ is the local slope of the linear power-spectrum at the typical 
wavenumber $k_s$ probed by the observations at redshift $z$. From 
eq.(\ref{kapk1}), we define this typical wavenumber as:
\beq
k_s(z) = \frac{1}{\De(z)\theta_s} .
\label{ks}
\eeq
Finally, we introduce the power, $\Delta^2(k)$, per logarithmic interval 
defined as:
\beq
\Delta^2(k,z) = 4 \pi k^3 P(k,z) .
\label{Delta2k}
\eeq
Then, for intermediate regimes defined by $1 < \Delta^2(k_s,z) < 
\Delta_{\rm vir}(z)$, we use the simple linear interpolation:
\beq
S_3(z)= S_3^{\rm QL} + \frac{\Delta^2(k_s,z) -1}{\Delta_{\rm vir}(z)-1} 
\left( S_3^{\rm NL} - S_3^{\rm QL} \right) .
\label{S32}
\eeq
For $\Delta^2(k_s,z)<1$ (i.e. the quasi-linear regime), we take 
$S_3=S_3^{\rm QL}$, while for $\Delta^2(k_s,z) > \Delta_{\rm vir}(z)$ 
(i.e., the highly non-linear regime), we take $S_3=S_3^{\rm NL}$.
Here $\Delta_{\rm vir}$ is the density contrast at virialization given by 
the usual spherical collapse (thus $\Delta_{\rm vir} \sim 178$ for a 
critical-density universe). Indeed, the threshold $\Delta^2 > 
\Delta_{\rm vir}$ describes the highly non-linear regime where most of the 
matter at scale $1/k$ has collapsed into non-linear structures.

Note that the function $\zeta(\tau)$ which we use as
a mere intermediate tool to parameterize the generating $\varphi(y)$ also
has a more physical meaning. As recalled in 
Sect.~\ref{Cumulant generating function for the density contrast}, in the
quasi-linear regime it can actually be derived from the equations of motion
and it is closely related to the spherical dynamics. On the other hand, in the
highly non-linear regime it can also be interpreted as an approximation to
the vertex generating function $\zeta_{\nu}(\tau)$ which appears within the
framework of ``minimal tree-models'', as we shall discuss below in 
Sect.~\ref{Minimal tree-model}. Since we use a constant skewness 
$S_3^{\rm NL}$in the highly non-linear regime this parameterization for 
$\varphi(y)$ is consistent with the stable-clustering {\it Ansatz}. However,
if needed it is straightforward to include deviations from this {\it Ansatz} by
incorporating some additional dependence on time into the skewness $S_3$.

Finally, we obtain the non-linear evolution of the power-spectrum from the
prescription given by Peacock \& Dodds (1996). This completes the description
of the pdf $\cP(\delta_R)$ and of the cumulants $\lag \delta_R^p\rag_c$ for
all scales and times and for any cosmological parameters.

\subsection{Minimal tree model}
\label{Minimal tree-model}

\begin{figure}
\protect\centerline{\epsfysize = 3 cm \epsfbox{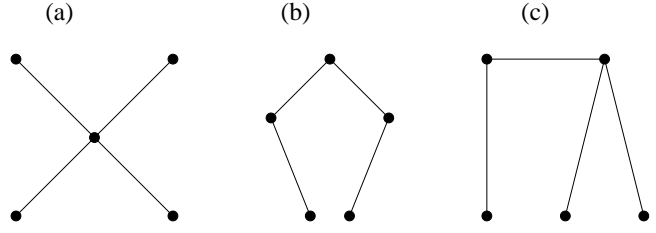}}
\caption{The various topologies one can build for the 5-point connected 
correlation function within the framework of a tree-model as in (\ref{tree}). 
Graph (a) is a ``stellar diagram'' (i.e., $(p-1)$ points are linked to a 
central point), while graph (b) is a ``snake diagram'' (i.e., one 
successively visits all points along one curve with only two end-points) 
and graph (c) is a ``hybrid diagram''.}
\end{figure}

The generating function $\varphi(y)$ introduced in 
Sect.~\ref{Cumulant generating function for the density contrast}
defines the properties of the density contrast $\delta_R$ smoothed over
spherical cells of radius $R$. In order to fully describe the density field
we actually need to specify the detailed behaviour of the many-body 
connected correlation functions, $\xi_p(\bx_1,..,\bx_p)$, defined by 
(Peebles, 1980):
\beq
\xi_p (\bx_1,..,\bx_p) = \lag \delta(\bx_1) .. \delta(\bx_p) \rag_c .
\label{xip}
\eeq
Indeed, the cumulants $\lag\delta_R^p\rag_c$ introduced in the previous 
section only measure the mean of these correlations over spherical cells:
\beq
p \geq 2 \; : \;\; \lag\delta_R^p\rag_c = \xib_p = \int_V 
\frac{\d\bx_1 .. \d\bx_p}{V^p} \; \xi_p (\bx_1,..,\bx_p) .
\label{xibp}
\eeq
Therefore, the cumulants $\lag\delta_R^p\rag_c$ are insufficient if we are
interested in real-space filters which are significantly different from a
spherical top-hat. As shown in Valageas (2000a, b), this is not very 
important for the convergence, $\kappa$, but it is a key point for the 
aperture-mass, $\Map$, see Bernardeau \& Valageas (2000), or the shear, 
$\gam$, see Valageas, Barber \& Munshi (2003), which involve compensated
filters. This leads one to introduce more precise models for the density
field which fully describe the correlations $\xi_p(\bx_1,..,\bx_p)$.

One such model is the ``minimal tree-model'' which is actually a specific
case of the more general ``tree-models''. The latter are defined by the 
hierarchical property (Schaeffer, 1984, and Groth \& Peebles, 1977):
\beq
\xi_p(\bx_1, .. ,\bx_p) = \sum_{(\alpha)} Q_p^{(\alpha)} \sum_{t_{\alpha}} 
\prod_{p-1} \xi_2(\bx_i,\bx_j) 
\label{tree}
\eeq
where $(\alpha)$ is a particular tree-topology connecting the $p$ points 
without making any loop, $Q_p^{(\alpha)}$ is a parameter associated with the 
order of the correlations and the topology involved, $t_{\alpha}$ is a 
particular labeling of the topology, $(\alpha)$, and the product is made over 
the $(p-1)$ links between the $p$ points with two-body correlation functions. 
We show in Fig.1, taken from Valageas, Barber \& Munshi (2003), the three 
topologies which appear within this framework for the 5-point connected 
correlation function.

Then, the minimal tree-model corresponds to the specific case where the 
weights $Q_p^{(\alpha)}$ are given by (Bernardeau \& Schaeffer, 1992): 
\beq
Q_p^{(\alpha)} = \prod_{\mbox{vertices of } (\alpha)} \nu_q
\label{mintree}
\eeq
where $\nu_q$ is a constant weight associated to a vertex of the tree 
topology with $q$ outgoing lines. The advantage of this minimal tree-model
is that it is well-suited to the computation of the cumulant generating
functions as defined in eq.(\ref{phidel2}) for the density contrast 
$\delta_R$. Indeed, for an arbitrary real-space filter, $F(\bx)$, which 
defines the random variable $\dum$ as:
\beq
\dum = \int \d\bx \; F(\bx) \; \delta(\bx) \hspace{0.4cm} \mbox{and} 
\hspace{0.4cm} \xidum= \lag \dum^2 \rag ,
\label{mu}
\eeq
it is possible to obtain a simple implicit expression for the 
generating function, $\phidum(y)$ (see Bernardeau \& Schaeffer, 1992,
and Jannink \& Des Cloiseaux, 1987): 
\beqa
{\displaystyle \phidum(y)} & = & {\displaystyle y \int \d\bx \; F(\bx) \; 
\left[ \zeta_{\nu}[ \tau(\bx)] - \frac{\tau(\bx) \zeta'_{\nu}[\tau(\bx)]}{2} 
\right] } 
\label{implicitphi} \\ 
{\displaystyle \tau(\bx) } & = & {\displaystyle -y 
\int \d\bx' \; F(\bx') \; \frac{\xi_2(\bx,\bx')}{\xidum} \; 
\zeta'_{\nu}[\tau(\bx')] } 
\label{implicittau}
\eeqa
where the function $\zeta_{\nu}(\tau)$ is defined as the generating function 
for the coefficients $\nu_p$:
\beq
\zeta_{\nu}(\tau) = \sum_{p=1}^{\infty} \frac{(-1)^p}{p!} \; \nu_p \; \tau^p 
\hspace{0.4cm} \mbox{with} \hspace{0.4cm} \nu_1 = 1.  
\label{zeta1}
\eeq
Of course, the generating function, $\phidum(y)$, depends on the filter, $F$, 
which defines the variable, $\dum$. If the real-space filter, $F(\bx)$, is
close to a top-hat (this is actually the case for the smoothed density 
contrast, $\delta_R$, defined in eq.(\ref{deltaR1}), where $F(\bx)$ is the
top-hat of radius $R$ normalised to unity), a simple ``mean field'' 
approximation which provides very good results 
(Bernardeau \& Schaeffer, 1992) is to integrate $\tau(\bx)$ over the 
relevant volume $V$ in eq.(\ref{implicittau}) with a weight $F(\bx)$ and 
then to approximate $\tau(\bx)$ by a constant $\tau$. This leads to the
simple system (\ref{phiMF})-(\ref{tauMF}) we have already encountered in
Section~\ref{Cumulant generating function for the density contrast}.
Therefore, in case the density field is described by such a minimal 
tree-model, the function $\zeta(\tau)$ introduced in 
Section~\ref{Cumulant generating function for the density contrast} to
parameterize the cumulant generating function $\varphi(y)$ can also be
interpreted as a good approximation to the generating function 
$\zeta_{\nu}(\tau)$ of the vertices $\nu_p$.

\subsection{Stellar model}
\label{Stellar model}

In a previous work (Valageas, Barber \& Munshi, 2003) where we studied 
the properties of the shear, $\gam$, we introduced another simple model
for the many-body correlations of the density field which is well-suited
to the computation of weak-lensing effects. This ``stellar-model'' is
another case of the tree-models defined in (\ref{tree}), where we only keep 
the stellar diagrams (e.g., the graph (a) in Fig.1 for the 5-point connected 
correlation). Thus, the $p-$point connected correlation $\xi_p$ of the 
density field can now be written as:
\beq
\xi_p(\bx_1, .. ,\bx_p) = \frac{\tS_p}{p} \; \sum_{i=1}^p \prod_{j\neq i} 
\xi_2(\bx_i,\bx_j) .
\label{xistar1}
\eeq
The advantage of the stellar-model (\ref{xistar1}) is that it leads to 
very simple calculations in Fourier space. Indeed, eq.(\ref{xistar1}) reads 
in Fourier space:
\beq
\lag \delta(\bk_1) .. \delta(\bk_p) \rag_c = \frac{\tS_p}{p} \; 
\delta_D(\bk_1+..+\bk_p) \sum_{i=1}^p \prod_{j\neq i} P(k_j) .
\label{corrstar1}
\eeq
Of course, the Dirac factor, $\delta_D(\bk_1+..+\bk_p)$, simply translates 
the fact that the many-body correlations are invariant through translations. 
The coefficients $S_p$ and $\tS_p$ are related by:
\beq
S_p = \tS_p \int_0^1 dt \; 3t^2 \; \left[ \frac{\int \frac{dk}{k} \; 
\Delta^2(k) F(kR) \frac{\sin(kRt)}{kRt}}{\int \frac{dk}{k} \; \Delta^2(k) 
F^2(kR)} \right]^{p-1}
\label{SptSp1}
\eeq
where we introduced the Fourier transform, $F(kR)$, of a 3-d top-hat of 
radius $R$:
\beq
F(kR) = \int \frac{\d\bx}{V} \; e^{i\bk.\bx} = 3 \; 
\frac{\sin(kR) - (kR) \cos(kR)}{(kR)^3} .
\eeq
However, in the following we shall use the simple approximation:
\beq
\tS_p \simeq S_p .
\label{SptSp2}
\eeq
Alternatively, we may define the function $\varphi(y)$ obtained from 
(\ref{phiMF})-(\ref{tauMF}) and our choice of $\zeta(\tau)$ as the 
generating function of the coefficients $\tS_p$, rather than $S_p$, through 
its Taylor expansion at $y=0$.

\subsection{Comments regarding our prescription versus the stable-clustering 
{\it Ansatz} and the halo model}
\label{Comments}

Here we must point out that our prescription where the
$p-$point correlations are expressed from a tree-model but the coefficients
$S_p$ (whence $ Q_p^{(\alpha)}$ in eq.(\ref{tree}))
are allowed to depend on scale and time is not a standard hierarchical 
model.\footnote{We would like to thank A. Taruya for calling our attention
to the risks of confusion with more standard models.}
This is only a simple model which yields the angular dependence of the
many-body correlations, over a given scale, keeping the freedom to vary their
amplitude with scale and time. Therefore, it is mainly a phenomenological tool
which is not exactly self-consistent. Indeed, if we are given the two-point
correlation and the coefficients $S_p$ at a given scale, applying the stellar 
model to all scales would only be fully self-consistent with constant $S_p$.
Otherwise, there is some ambiguity: given a configuration for the $p-$point
correlation, which scale should be attached to this configuration so as to
define the relevant parameters $S_p$? For our purposes, our prescription
is sufficient and well-defined because smoothed weak-lensing observables like
the smoothed convergence $\kappa_s$ precisely select a ``typical'' length 
scale at each redshift along the line of sight: the radius of the
smoothing cone $\De(z)\theta_s$. In Fourier space, this is given by the
typical wavenumber which we defined in eq.(\ref{ks}). Of course there remains
a small ambiguity since we could as well have chosen $1.5/(\De\theta_s)$
but the associated uncertainty is smaller than the accuracy of our 
analytical model.

The reason why we use such a two-step procedure is simply that we know the
coefficients $S_p$ (such as the skewness $S_3$) do depend on time and scale
in CDM cosmologies. Therefore, it is necessary to keep such a freedom for
the generating function $\varphi(y)$, as described in 
Sect.~\ref{A simple parameterization for the generating function}. Then,
as discussed below eq.(\ref{xibp}), it happens that for some purposes the
coefficients $S_p$ are not sufficient as one needs the angular dependence
of the many-body correlations over some typical scale. This is why
we introduced the two tree models presented in Sect.~\ref{Minimal tree-model}
and Sect.~\ref{Stellar model}, which are understood to apply around some
typical length scale as we explained above.

As recalled in 
Sect.~\ref{A simple parameterization for the generating function},
our prescription in the highly non-linear regime for $\varphi(y)$ is 
consistent with the 
stable-clustering {\it Ansatz} (Peebles 1980). However, some results from
numerical simulations (e.g., Smith et al., 2002) suggest that the 
stable-clustering {\em Ansatz} is only approximate as continuing mergers
yield deviations from this simple model. Such departures could be included
within our approach through the non-linear power-spectrum, $P(k)$, and through
an additional dependence on redshift and scale for the skewness, $S_3$, in 
the non-linear regime. If required, one could simply choose another 
parameterization for the cumulant generating function $\varphi(y)$ which
would involve a specific dependence on redshift and scale obtained from a fit
to numerical simulations or from some other model for the density field.
Indeed, our calculations do not depend on the parameterization 
(\ref{phiMF})-(\ref{S32}) and we simply express the generating function 
$\phikap(y)$ of the convergence $\kaphs$ in terms of the generating function
$\varphi(y)$ of the density contrast. Therefore, one simply needs to replace
this function $\varphi(y;k_s,z)$ by one's specific choice for the dependence
on $k_s$ and $z$. 

In particular, an alternative to the stable-clustering {\it Ansatz} is 
provided by the ``halo model'' where the density field is described 
through a random distribution of dark matter halos, modulated by the 
large-scale matter distribution (e.g., Seljak 2000, Peacock \& Smith 2000). 
This allows one to include deviations from the stable-clustering {\it Ansatz} 
brought by the mergings and disruptions of these halos. 
However, as shown in Valageas (1999), note that the simplest halo 
model where all halos would have the same density profile (rescaled to their
virial radius) is strongly inconsistent with numerical simulations as it 
yields $\xib_p(R) \sim R^{-(p\beta-3)}$ where $\beta$ is the slope of the
inner density profile of these halos (i.e. $\rho_{\rm halo}(r) \propto 
r^{-\beta}$). Thus, the usual halo models involve a density profile which
depends on the mass of the halo, through a concentration parameter. As noticed
by Navarro et al. (1996) a good estimate is obtained by requiring the density
within the core radius to scale as the density of the universe at the redshift
when this mass scale turned non-linear. This can actually be seen as a way
to include some features of the stable-clustering {\it Ansatz} within the
halo model. Then, using such a dependence on $M$ and $n$ for the halo profiles
(or a fit from simulations) one can match the observed  non-linear 
power-spectrum (e.g., Smith et al. 2002) and possibly reproduce
higher order correlations. Indeed, as noticed in Valageas (1999), if the 
low-mass tail of the halo mass function shows a power-law behaviour of the form
$\eta(M) \d M/M \propto M^{\theta-1} \d M/M$ one recovers the 
stable-clustering {\it Ansatz} for $\theta \rightarrow 0$ (in which case
$\eta(M)$ also counts substructures within halos) so that one can probably 
obtain good results for higher order correlations by choosing a slope 
$\theta \ga 0$. Then, one could compute from such a model the cumulants
$\lag \delta_R^p\rag_c$ whence the generating function $\varphi(y)$. Next, one
can use all methods presented in this paper with this new function 
$\varphi(y)$.

However, it is not clear whether such a halo model can be made 
fully self-consistent since higher order correlations may be increasingly 
influenced by the substructures present within halos (see also Valageas 1999).
Moreover, this model cannot reproduce the behaviour of the density field
over low-density regions like voids and filaments (i.e. outside virialized 
halos). This prevents the computation of the full pdfs $\cP(\delta_R)$
and $\cP(\kappa)$ for $\delta_R \la 0$ and $\kappa \la 0$. 

Therefore,
in this work we shall only consider the simple parameterization 
(\ref{zetadef})-(\ref{S32}) described in 
Sect.~\ref{A simple parameterization for the generating function}.
Indeed, detailed comparisons with numerical simulations, presented below in 
Sect.~\ref{Results}, show that it already provides good predictions for 
the pdf $\cP(\kappa)$ of the convergence. Moreover, this model has the 
advantage of a great simplicity and it automatically shows the right 
behaviour on quasi-linear scales.

\section{Computation of the pdfs}
\label{Computation of the pdfs}

We now compute the pdf $\cP(\kappa)$ of the convergence following the method
developed in Valageas (2000a, b). From eq.(\ref{kap1}) or eq.(\ref{kapk1})
we express the cumulants of the smoothed normalised convergence 
$\lag \kaphs^p\rag_c$ in terms
of the many-body correlations of the density contrast. Next, after 
resummation of this series of cumulants we obtain the generating function
$\phikap$ of the convergence as in eq.(\ref{phidel2}) which yields the
pdf $\cP(\kaphs)$. Since this method has already been used in previous
works for the convergence (Valageas 2000a, b), the aperture-mass 
(Bernardeau \& Valageas 2000) and the shear (Valageas, Barber \& Munshi 2003)
we only briefly recall the main steps of this derivation, in order to
show where the various approximations one can introduce within this framework
come into play.

We first present in 
Sect.~\ref{Spherical approximation} and Sect.~\ref{Zeroth-order approximation}
two simple approximations which allow us to compute $\cP(\kappa)$ without any
assumption about the properties of the density field. They provide two simple
expressions for $\phikap(y)$ in terms of $\varphi(y)$ which can be used with
our parameterization described in 
Sect.~\ref{A simple parameterization for the generating function} or with any
alternative prescription for $\varphi(y)$ or $\cP(\delta_R)$. This allows
us to consider in Sect.~\ref{Lognormal approximation} a log-normal model for
$\cP(\delta_R)$. Finally, in Sect.~\ref{Minimal tree-model approximation} and
Sect.~\ref{Stellar approximation} we present exact calculations within the
framework of both the minimal tree-model and the stellar-model introduced in
Sect.~\ref{Minimal tree-model} and Sect.~\ref{Stellar model}, in order to
check in two explicit cases the model-independent approximations described
in Sect.~\ref{Spherical approximation} and 
Sect.~\ref{Zeroth-order approximation}.

\subsection{Spherical-cell approximation}
\label{Spherical approximation}

Following Valageas (2000a, b), see also Bernardeau \& Valageas (2000), 
we obtain from eq.(\ref{kap1}) for the cumulant of order $p$ of the 
smoothed normalised convergence $\kaphs$:
\beqa
\lag \kaphs^p\rag_c & = & \int_0^{\chi_s} \d\chi \; \wh^p 
\int_{-\infty}^{\infty} \prod_{i=2}^{p} \d\chi_i \int_0^{\theta_s} 
\prod_{i=1}^{p} \frac{\d{\vec \vartheta}_i}{\pi \theta_s^2} \nonumber \\ 
& & \times \; \xi_p\left( \bea{l} 0 \\ \De {\vec \vartheta}_1 \ea ,
\bea{l} \chi_2 \\ \De {\vec \vartheta}_2 \ea , .. , 
\bea{l} \chi_p \\ \De {\vec \vartheta}_p \ea ; z \right) .
\label{cum1}
\eeqa
Here we used the fact that the correlation length (beyond which the 
many-body correlations are negligible) is much smaller than the Hubble 
scale, $c/H(z)$ (where $H(z)$ is the Hubble constant at redshift $z$). 
Although the points $(\chi_i,\De {\vec \vartheta}_i)$ cover a cylinder 
of radius $\De \theta_s$ and length $L$ (with $L \rightarrow \infty$ 
in eq.(\ref{cum1})) rather than a sphere, we may approximate the integral 
over the $p-$point correlation $\xi_p$ as:
\beq
\int_{-\infty}^{\infty} \prod_{i=2}^{p} \d\chi_i \int_0^{\theta_s} 
\prod_{i=1}^{p} \frac{\d{\vec \vartheta}_i}{\pi \theta_s^2} 
\; \xi_p 
\simeq S_p \; \Ikap^{p-1},
\label{spher1}
\eeq
in a fashion similar to $\xib_p= S_p \xib_2^{\;p-1}$, see eqs.(\ref{phidel2}),
(\ref{xibp}). Here the coefficients $S_p(z)$ are evaluated at the wavenumber
$k_s$ defined in eq.(\ref{ks}), associated with the radius $\De(z)\theta_s$ of
the cylinder at redshift $z$, and we defined:
\beq
\Ikap(z)= \int_{-\infty}^{\infty} \d\chi' \int_0^{\theta_s} 
\frac{\d{\vec \vartheta}_1}{\pi \theta_s^2} 
\frac{\d{\vec \vartheta}_2}{\pi \theta_s^2}
\; \xi_2 \left( \bea{l} 0 \\ \De {\vec \vartheta}_1 \ea ,
\bea{l} \chi' \\ \De {\vec \vartheta}_2 \ea ; z \right) .
\label{Ikap1}
\eeq
This quantity can also be written as (Valageas 2000b):
\beq
\Ikap(z)= \pi \int_0^{\infty} \frac{\d k}{k} \; \frac{\Delta^2(k,z)}{k} \; 
W(k \De \theta_s)^2 ,
\label{Ikap2}
\eeq
where the filter $W$ was defined in eq.(\ref{Wk}). Thus, 
using eq.(\ref{spher1}) we can write the cumulants (\ref{cum1}) as:
\beq
\lag \kaphs^p\rag_c = \int_0^{\chi_s} \d\chi \; \wh^p \; S_p \; \Ikap^{p-1} .
\label{cum2}
\eeq
Then, using the expression (\ref{phidel2}) for the cumulant generating 
function $\phikap(y)$ associated with the smoothed normalised convergence
$\kaphs$ we obtain:
\beq
\phikap(y) = \int_0^{\chi_s} \d\chi \; \frac{\xikap}{\Ikap} \;
\varphi\left( y \wh \frac{\Ikap}{\xikap} ; z \right) ,
\label{phispher1}
\eeq
where we used the resummation (\ref{phidel2}) for the coefficients $S_p$ and
we introduced the variance:
\beq
\xikap = \lag \kaphs^2\rag = \int_0^{\chi_s} \d\chi \; \wh^2 \; \Ikap .
\label{xikap1}
\eeq
Of course, eq.(\ref{xikap1}) is exact, within the small-angle approximation,
and the pdf $\cP(\kaphs)$ is finally obtained from the inverse Laplace 
transform (\ref{Pdel1}):
\beq
\cP(\kaphs) = \inta \frac{\d y}{2\pi i \xikap} \; 
e^{[\kaphs y - \phikap(y)] /\xikap} .
\label{Pkap1}
\eeq
We shall refer to the approximation (\ref{spher1})-(\ref{phispher1}) as 
the ``spherical-cell approximation'' because it is based on the use for 
the average of the many-body correlations over a cylinder (\ref{spher1}) 
of their average over a spherical cell:
\beq
\frac{\lag \xi_p \rag_{\rm cyl.}}{\lag \xi_2 \rag_{\rm cyl.}^{\;p-1}} \simeq
\frac{\lag \xi_p \rag_{\rm spher.}}{\lag \xi_2 \rag_{\rm spher.}^{\;p-1}}
= S_p .
\label{spher2}
\eeq
In this equation only the notation $\lag .. \rag$ is a spatial average, over
spherical cells or cylinders. This approximation should not be confused 
with an approximation based on some ``spherical
dynamics''. It merely assumes that the dependence on geometry of the ratios
$\xib_p/\xib_2^{\;p-1}$ can be neglected. The advantage of this approximation
is that it can be applied to any model for the density field. Indeed, it does
not involve the detailed behaviour of the many-body correlations 
$\xi_p (\bx_1,..,\bx_p)$: we only need the pdf $\cP(\delta_R)$ (or the
associated generating function $\varphi(y)$) of the density contrast within
spherical cells. In particular, we shall see in the following sections that
eq.(\ref{phispher1}) can be recovered from more specific models. As recalled
in Sect.~\ref{Minimal tree-model}, the drawback of this approximation is
that it cannot be extended to compensated filters like those used for the
aperture-mass or the shear.

\subsection{Mean-redshift approximation}
\label{Zeroth-order approximation}

We see in eq.(\ref{phispher1}) that the projection of the 3-d density field
onto the 2-d convergence is described by a simple integration along the line 
of sight of the generating function $\varphi(y)$ (within the framework of the
spherical-cell approximation). We wrote explicitly in eq.(\ref{phispher1})
the dependence on redshift of $\varphi(y)$ along the line of sight, which 
follows the evolution of the wavenumber probed at redshift $z$, see 
eq.(\ref{ks}), as well as the growth of density fluctuations from the linear
to non-linear regime, see eq.(\ref{S32}). As noticed in Valageas (2000a, b),
we may approximate this integral along the line of sight by a mean value.
Since both $\chi_s \xikap/\Ikap$ and $\wh \Ikap/\xikap$ are typically 
close to unity one may simply use:
\beq
\phikap(y) \simeq \varphi(y;z_*) ,
\label{zero1}
\eeq
which still obeys the constraint $\phikap(y) = -y^2/2 +...$ at $y=0$, see
eq.(\ref{phidel2}). In eq.(\ref{zero1}), we take for the generating function 
$\varphi(y)$ its value at the typical redshift $z_*$ (and wavenumber 
$k_s(z_*)$) which we define as the location of the maximum of the selection 
function $\wh(z,z_s)$. This yields $z_* \sim z_s/2$.
We shall refer to this approximation (\ref{zero1}) as the
``mean-redshift approximation'' because it replaces the projection onto
two dimensions by a typical value along the line of sight. Eq.(\ref{zero1}) 
actually means that the pdf $\cP(\kaphs)$ of the smoothed normalised
convergence is directly given by the pdf $\cP(\delta_R)$ of the density 
contrast at the typical scale $1/k_s(z_*)$ and redshift $z_*$ probed by the 
observation:
\beq
\cP(\kaphs) \simeq \cP(\delta_R \rightarrow \kaphs , \xib_2 \rightarrow 
\xikap ; k_s(z_*),z_*) .
\label{zero2}
\eeq
Eq.(\ref{zero2}) clearly shows that the statistical properties of the
convergence provide an efficient probe of the density field. In particular,
we can hope to measure the departures from Gaussianity brought by the
non-linear gravitational dynamics from $\cP(\kaphs)$.

\subsection{Log-normal approximation}
\label{Lognormal approximation}

It is clear that both the spherical-cell approximation and the 
mean-redshift approximation presented in 
Sect.~\ref{Spherical approximation} and Sect.~\ref{Zeroth-order approximation}
are model-independent. Indeed, they do not assume any specific behaviour for
the generating function $\varphi(y)$ or the many-body correlations $\xi_p$.
However, the accuracy of these simple approximations may depend on the
properties of the density field. For instance, if $\varphi(y)$ or the
selection function $\wh$ strongly evolve with redshift the average used for
$\varphi(y)$ within the mean-redshift approximation (\ref{zero1})
may be too inaccurate. Nevertheless, within the framework of the 
mean-redshift approximation 
(\ref{zero1})-(\ref{zero2}) one may directly use any model for the pdf
of the density contrast to estimate $\cP(\kaphs)$, with no need to
compute the generating function $\varphi(y)$ itself. 
Thus, Taruya et al. (2002) used this mean-redshift approximation 
to compute the pdf $\cP(\kaphs)$ from a log-normal
distribution for the density contrast. That is, they used for $\cP(\delta_R)$
in eq.(\ref{zero2}) the expression:
\beqa
\cP_{\rm ln}(\delta_R) & = &
\frac{1}{(1+\delta_R)\sqrt{2\pi\ln(1+\xib_2)}} \nonumber \\
& & \times \; 
\exp\left( - \frac{\ln^2[(1+\delta_R)\sqrt{1+\xib_2}]}{2\ln(1+\xib_2)} 
\right) .
\label{ln1}
\eeqa
As noticed by Taruya et al. (2002) the log-normal pdf
is a simple example which violates the stable-clustering {\it Ansatz}.
However, as we explained in Sect.~\ref{Comments} our method does not
need the stable-clustering {\it Ansatz} to be valid so that the log-normal pdf
is fully consistent with the approach recalled in 
Sect.~\ref{Spherical approximation} and with eqs.(\ref{zero1})-(\ref{zero2}).
One advantage of this model over the use of the generating function
$\varphi(y)$ obtained from eqs.(\ref{phiMF})-(\ref{tauMF}) is that one does
not need to compute the inverse Laplace transform (\ref{Pdel1}) and there is
one fewer parameter, indeed the log-normal pdf (\ref{ln1}) only involves
the variance $\xib_2$. On the other hand, we can expect the additional 
parameter $\kappa$ (or $S_3$) which enters our parameterization 
(\ref{zetadef}) and allows us to take into account the dependence of the
skewness on the slope of the power-spectrum to improve the accuracy of this
prescription over the log-normal approximation. 

For instance, as noticed
in Bernardeau (1994), in the quasi-linear regime the log-normal approximation
is quite good for $n=-1$ but worsens for different power-spectra. We may note
here that in the quasi-linear limit our prescription 
(\ref{phiMF})-(\ref{zetadef}) actually yields back the log-normal pdf for
$S_3 \rightarrow 3$, where $\zeta(\tau) \rightarrow e^{-\tau}-1$ (e.g., 
App.A of Valageas, 2002). From eq.(\ref{S31}) this corresponds to $n=-1.14$
which explains why the log-normal pdf agrees with numerical simulations for
$n \simeq -1$ in the quasi-linear regime. However, we can expect significant
discrepancies in other cases where $S_3$ is much larger than $3$ (e.g., for
$n \la -2$ or in the highly non-linear regime).

\subsection{Minimal tree-model approximation}
\label{Minimal tree-model approximation}

In the spherical-cell approximation presented in 
Sect.~\ref{Spherical approximation} we estimated the average over cylinders
of the correlations $\xi_p$ by the simple approximation (\ref{spher1}). As
discussed above, the advantage of this approach is that it does not require
much information about the behaviour of the density field. However, one may
wonder what is the actual accuracy of this simple estimate. To tackle this
point we need to evaluate exactly the l.h.s. in eq.(\ref{spher1}) for some
specific models and compare the result with the r.h.s. approximation. We 
could also compute both quantities directly in N-body simulations. Note that
the pdf $\cP(\kaphs)$ measured in numerical simulations does not allow us to
test the approximation (\ref{spher1}) itself since it also involves the
parameterization used for the generating function $\varphi(y)$.
Therefore, it is interesting to compute the pdf 
$\cP(\kaphs)$ in a more rigorous way, without using eq.(\ref{spher1}), for
some specific cases.

Obviously, to do so, we need the detailed behaviour of the $p-$point 
correlation $\xi_p(\bx_1,..,\bx_p)$. One simple example where we can sum up
the cumulants $\lag \kaphs^p\rag_c$ so as to derive the generating function
$\phikap(y)$ is the minimal tree-model recalled in 
Sect.~\ref{Minimal tree-model}. This case has already been studied in
Valageas (2000b) and Bernardeau \& Valageas (2000). Here we shall simply 
recall how we can recover the spherical-cell approximation (\ref{spher1}) 
within this framework. As seen from eq.(\ref{cum1}), in order to make 
some progress we need to evaluate the quantities:
\beqa
\om_p({\vec \vartheta}_1,..,{\vec \vartheta}_p;z) & = & 
\int_{-\infty}^{\infty} \prod_{i=2}^{p} \d\chi_i \; 
\xi_p\left( \bea{l} 0 \\ \De {\vec \vartheta}_1 \ea , .. ,
\bea{l} \chi_p \\ \De {\vec \vartheta}_p \ea ; z \right) \nonumber \\
\label{om1}
\eeqa
Then, as noticed in Valageas (2000b), if the 3-d correlations $\xi_p$ obey
a tree-model as in eq.(\ref{tree}) the 2-d correlations $\om_p$ exhibit
the same tree-structure:
\beq
\om_p({\vec \vartheta}_1,..,{\vec \vartheta}_p;z) = \sum_{(\alpha)} 
Q_p^{(\alpha)} \sum_{t_{\alpha}} \prod_{p-1} 
\om_2({\vec \vartheta}_i,{\vec \vartheta}_j;z) 
\label{omtree1}
\eeq
with:
\beq
\om_2({\vec \vartheta}_1,{\vec \vartheta}_2;z) = \pi \int_0^{\infty} 
\frac{\d k}{k} \; \frac{\Delta^2(k,z)}{k} \; 
J_0(k \De |{\vec \vartheta}_1 - {\vec \vartheta}_2|) ,
\label{om21}
\eeq
where $J_0$ is the Bessel function of order 0. Next, in the case of a minimal
tree-model (\ref{mintree}) we can perform the resummation 
(\ref{implicitphi})-(\ref{implicittau}) for the 2-d correlations $\om_p$, since
the latter obey the same minimal tree-model from eq.(\ref{omtree1}). This 
yields (see Bernardeau \& Valageas 2000 for details):
\beq
\phikap(y) = \int_0^{\chi_s} \d\chi \; \frac{\xikap}{\Ikap} \;
\varphi_{\rm cyl.}\left( y \wh \frac{\Ikap}{\xikap} ; z \right) ,
\label{phitree1}
\eeq
where we introduced the 2-d generating function $\varphi_{\rm cyl.}$ 
associated with the 2-d correlations $\om_p$, given by the resummation:
\beqa
{\displaystyle \varphi_{\rm cyl.}(y)} & = & {\displaystyle y 
\int_0^{\theta_s} \frac{\d{\vec \vartheta}}{\pi \theta_s^2} \; 
\left[ \zeta_{\nu}[ \tau({\vec \vartheta})] 
- \frac{\tau({\vec \vartheta}) \zeta'_{\nu}[\tau({\vec \vartheta})]}{2} 
\right] } 
\label{phiom} \\ 
{\displaystyle \tau({\vec \vartheta}) } & = & {\displaystyle -y 
\int_0^{\theta_s} \frac{\d{\vec \vartheta}'}{\pi \theta_s^2} \; 
\frac{\om_2({\vec \vartheta},{\vec \vartheta}';z)}{\omb_2(z)} \; 
\zeta'_{\nu}[\tau({\vec \vartheta}')] } 
\label{tauom}
\eeqa
Here we introduced the angular average $\omb_2$ of the 2-d correlation 
$\om_2$, which is actually equal to the quantity $\Ikap$ defined in 
eqs.(\ref{Ikap1})-(\ref{Ikap2}):
\beq
\omb_2(z) = \int_0^{\theta_s} \frac{\d{\vec \vartheta}_1}{\pi \theta_s^2}
\frac{\d{\vec \vartheta}_2}{\pi \theta_s^2} \;
\om_2({\vec \vartheta}_1,{\vec \vartheta}_2;z) = \Ikap(z) ,
\label{omb1}
\eeq
and we substituted the notation $\Ikap$ to $\omb_2$ into eq.(\ref{phitree1}).
Note the similarity of eq.(\ref{phitree1}) with eq.(\ref{phispher1}) obtained
from the spherical approximation. Moreover, we see that for a minimal 
tree-model the projection from three dimensions onto two dimensions does not
only yield the integration along the line of sight apparent in both 
eq.(\ref{phitree1}) and eq.(\ref{phispher1}). It also entails the change from
$\varphi(y)$, associated with 3-d spherical cells, to $\varphi_{\rm cyl.}$,
associated with a 2-d top-hat (after we integrated along the 
longitudinal direction, 
see eq.(\ref{om1})). Thus, this projection effect was actually neglected by
the spherical-cell approximation (\ref{spher1}). However, if we use the 
mean-field approximation discussed below eq.(\ref{zeta1}) (i.e. $\tau(\bx)$ is 
approximated by a constant over the relevant volume of integration), both 
for 3-d spherical cells (which amounts to the approximation 
$\zeta(\tau)\simeq\zeta_{\nu}(\tau)$) and for the 2-d top-hat of angular 
radius $\theta_s$ which appears in eqs.(\ref{phiom})-(\ref{tauom}), we obtain
$\varphi_{\rm cyl.}(y)=\varphi(y)$ and we recover eq.(\ref{phispher1}).
Therefore, within the framework of a minimal tree-model for the many-body
correlations the spherical approximation presented in 
Sect.~\ref{Spherical approximation} can be interpreted as the usual 
mean-field approximation. This is not surprising since the latter 
approximation actually neglects the geometry of the filter $F(\bx)$ associated
with any random variable $s$, see eqs.(\ref{mu})-(\ref{implicittau}), so
that the coefficients $S_p$ must be the same for spherical cells and
cylinders.

The mean-field approximation has already been shown to provide very good
results in the case of the 3-d top-hat (Bernardeau \& Schaeffer 1992) hence
we can expect a similar accuracy for the 2-d top-hat. This suggests that the
spherical-cell approximation (\ref{spher1}) should be quite accurate for most
models of the density field, even beyond the class of minimal tree-models.

\subsection{Stellar-model approximation}
\label{Stellar approximation}

Finally, we consider the second explicit model presented in 
Sect.~\ref{Analytical models for the density field}: the stellar model 
introduced in Sect.~\ref{Stellar model}. The calculation of the generating
function $\phikap(y)$ can again be performed explicitly and we can also
recover the result (\ref{phispher1}) obtained from the spherical-cell 
approximation through another mean field approximation. Moreover, we shall
compare the exact results obtained from this approach with those of the
spherical-cell approximation and we shall check in the Figures that both
predictions are indeed very close, in agreement with the discussions above.
The stellar model (\ref{xistar1})-(\ref{corrstar1}) was already introduced
in Valageas, Barber \& Munshi (2003) to compute the pdf of the smoothed shear 
components, $\gam_{1s}$ and $\gam_{2s}$, as well as the pdf of the 
smoothed modulus, $\gam_s=|\gam_{1s}+i\gam_{2s}|$, hence we shall only recall
here the main steps of the calculation since it proceeds in the same fashion
for the smoothed convergence, $\kappa_s$, which we study in this paper.
Within this framework it is convenient to work in Fourier space, hence we
write the cumulant $\lag \kaphs^p\rag_c$ as:
\beqa
\lefteqn{ \lag \kaphs^p \rag_c = \int_0^{\chi_s} \d\chi_1 \wh_1^p 
\int_{-\infty}^{\infty} \prod_{i=2}^{p} \d\chi_i \int \prod_{j=1}^{p} 
\d\bk_j \; W(k_{\perp j} \De \theta_s) } \nonumber \\ 
& & \times \left( \prod_{l=1}^{p} e^{i k_{\parallel l} \chi_l} \right) \; 
\tS_p \; \delta_D(\bk_1+..+\bk_p) P(k_2) .. P(k_p) ,
\label{cumk1}
\eeqa
where we used eq.(\ref{kapk1}) and eq.(\ref{corrstar1}), and the fact that 
the correlation length is much smaller than the Hubble scale. Thus, 
eq.(\ref{cumk1}) is equivalent to eq.(\ref{cum1}) which was written in 
real space, where we used the stellar model (\ref{corrstar1}) for the
correlation $\lag \delta(\bk_1) .. \delta(\bk_p) \rag_c$. Next, using the 
usual small-angle approximation (i.e. $P(k_j) \simeq P(k_{\perp j})$), we 
can perform the integration over $\chi_2,..,\chi_p$ and 
$k_{\parallel 1},..,k_{\parallel p}$, which yields:
\beqa
\lag \kaphs^p \rag_c & = & \int \frac{\d\chi}{2\pi} (2\pi\wh)^p 
\int \prod_{j=1}^{p} \d\bk_{\perp j} \; W(k_{\perp j} \De \theta_s) 
\nonumber \\ 
& & \times \;\tS_p \; \delta_D(\bk_{\perp 1}+..+\bk_{\perp p}) 
P(k_{\perp 2}) .. P(k_{\perp p}) .
\label{cumk2}
\eeqa
Then, using the standard exponential representation of the Dirac distribution
(see Valageas, Barber \& Munshi 2003), integrating over the angles of the
transverse wavenumbers $\bk_{\perp 1},..,\bk_{\perp p}$ and over 
$|\bk_{\perp 1}|$, we obtain:
\beq
\lag \kaphs^p \rag_c = \int_0^{\chi_s} \d\chi \; \wh^p \int_0^1 \d t
\; 2t \; \tS_p \; \Istel^{p-1} ,
\label{cumk3}
\eeq
where we introduced the quantity:
\beq
\Istel(t,z) = \pi \int_0^{\infty} \frac{\d k}{k} \; 
\frac{\Delta^2(k,z)}{k} \; W(k \De \theta_s) \; J_0(t k \De \theta_s) .
\label{Istel1}
\eeq
We used an asterix ``*'' in the notation $\Istel(t,z)$ in order to 
distinguish this quantity from $\Ikap(z)$ introduced in 
eqs.(\ref{Ikap1})-(\ref{Ikap2}) and to recall that $\Istel$ appears within
the stellar model. Finally, we can resum the cumulants 
$\lag \kaphs^p \rag_c$, using the resummation (\ref{phidel2}) where we replace
$S_p$ by $\tS_p$ (as discussed in Sect.~\ref{Stellar model} this may also
be taken as the definition of $\varphi(y)$), and we get:
\beq
\phikap(y) = \int_0^{\chi_s} \d\chi \int_0^1 \d t \; 2t \; 
\frac{\xikap}{\Istel} \; \varphi\left( y\wh \frac{\Istel}{\xikap};z \right) .
\label{phistel1}
\eeq
Thus, we again obtain a result very similar to the spherical-cell approximation
(\ref{phispher1}). As for the minimal tree-model, we see through this second 
exact calculation that the projection effects also lead to a modification of 
the 2-d generating function associated with a given redshift $z$ as compared 
with $\varphi(y)$. For the minimal tree-model, this led to the 2-d 
generating function $\varphi_{\rm cyl.}(y)$ obtained in 
eqs.(\ref{phiom})-(\ref{tauom}). For the stellar model, this projection effect
is described by the integration over $t$ of the generating function 
$\varphi$ which appears in the r.h.s. in eq.(\ref{phistel1}). Next, we can 
again recover the spherical-cell approximation by using
a suitable average for the integration over $t$. Indeed, noticing that
$\int_0^1 \d t \; 2t=1$ and $\int_0^1 \d t \; 2t \Istel=\Ikap$, we may
use the approximation:
\beq
\int_0^1 \d t \; 2t \; \frac{\xikap}{\Istel} \; 
\varphi\left( y \wh \frac{\Istel}{\xikap} ; z \right) \simeq 
\frac{\xikap}{\Ikap} \; \varphi\left( y \wh \frac{\Ikap}{\xikap} ; z \right) ,
\label{stelspher1}
\eeq
which yields back the spherical-cell result (\ref{phispher1}). We shall check
through numerical computations that the approximation (\ref{stelspher1})
is indeed very good, which gives a further justification for the simple
spherical-cell approximation (\ref{phispher1}).

\section{Convergence statistics from numerical simulations}
\label{Numerical simulations}

Our results for the pdfs of the convergence and the higher-order
moments established from the above analytical models are compared in
this paper with our results obtained from numerical simulations. The
numerical method we have used was originally devised by Couchman et
al. (1999) who developed a code for computing the 3-dimensional shear
matrices at locations within the simulation volumes. We have applied
this code to the simulations of the Hydra
Consortium\footnote{(http://hydra.mcmaster.ca/hydra/index.html)}
produced using the `Hydra' $N$-body hydrodynamics code (Couchman,
Thomas \& Pearce, 1995).

In this pilot study, in which analytical predictions are made and
compared with the results from numerical simulations for arbitrary
cosmological scenarios, we have chosen readily available simulation
data. Details of the two cosmological simulations used, LCDM and OCDM,
are summarised in Table 1, in which the angular size of the survey,
$\theta_{\rm survey}^2$, refers to the total angular size of the field
of view and corresponds to completely filling the front face of the
redshift 1 simulation volume. For source redshifts greater than 1, the
periodicity of the particle distributions was used to allow lines of
sight beyond the confines of the simulation volumes to be
included. This is unlikely to adversely affect the statistics for the
angular scales of interest, i.e., $1'.0$ to $8'.0$. In addition, there
is, of course, no affect for the source redshifts less than or equal
to 1.

Both cosmologies contained $86^3$ dark matter particles of mass $1.29
\times 10^{11}h^{-1}$ solar masses each, where $h$ expresses the value
of the Hubble parameter in units of 100~km~s$^{-1}$~Mpc$^{-1}$. A
variable particle softening, whose value reflected the density
environment of each particle, was used. The minimum value (in box
units) of the softening (for particles in the densest environments)
was $0.0007(1+z)$, where $z$ is the redshift of the particular
simulation volume. Each simulation volume had comoving side-dimensions
of 100$h^{-1}$Mpc and to avoid obvious structure correlations, each
was arbitrarily translated, rotated and reflected about each
coordinate axis.

The full procedure for specifying the coordinates for the
lines-of-sight, the locations within the simulations for the
computations of the shear and the procedure for the combination of the
3-d matrices to obtain the final Jacobian matrices is described in
detail by Barber (2002). In the present work, a total of $455 \times
455$ lines of sight were used and 300 regularly-spaced evaluation
locations were specified along each line of sight in each simulation
volume. In the LCDM cosmology, the full field of view was $2^{\circ}.6
\times 2^{\circ}.6$ and in the OCDM cosmology, $2^{\circ}.8 \times
2^{\circ}.8$.  The angular resolution in the LCDM cosmology was
$0'.34$, which equates to the minimum value of the particle softening
at the optimum redshift, $z = 0.36$, for lensing of sources at a
redshift of 1. In the case of the OCDM cosmology, the angular
resolution was $0'.37$.

\begin{table}
\begin{center}
\caption{The parameters used in the two cosmological
simulations. $\Gamma$ is the power spectrum shape parameter, $\Om$ is
the matter density parameter, $\Ol$ is the vacuum energy density
parameter, $\sigma_8$ is the normalisation on scales of $8h^{-1}$~Mpc,
$\theta_{\rm res}$ is the angular resolution and $\theta_{\rm
survey}^2$ is the angular size of the complete field throughout the
simulations.}
\label{tabsig2D}
\begin{tabular}{@{}lcccccc}
\hline
&$\Gamma$&$\Om$&$\Ol$&$\sigma_8$&$\theta_{\rm res}$&$\theta_{\rm survey}^2$\\
\hline
LCDM&0.25&0.3&0.7&1.22&$0'.34$&$2.6^{\circ} \times 2.6^{\circ}$ \\
OCDM&0.25&0.3&0.0&1.06&$0'.37$&$2.8^{\circ} \times 2.8^{\circ}$ \\
\hline
\end{tabular}
\end{center}
\end{table}

\begin{table*}
\begin{center}
\caption{The redshifts of the sources in each cosmology were selected
to give good statistical coverage throughout the overall range. In
each cosmology, they corresponded to the simulation box redshifts and
were chosen to be close to redshifts of 0.1, 0.2, 0.3, 0.4, 0.5, 0.6,
0.7, 0.8, 0.9, 1.0, 1.5, 2.0, 3.0 and 3.5. In this paper we refer to
the source redshifts as these approximate values, although in the
determination of our results, the actual redshift values, listed
below, were used.}
\label{tabsig2D}
\begin{tabular}{@{}lcccccccccccccc}
\hline
&$z_1$&$z_2$&$z_3$&$z_4$&$z_5$&$z_6$&$z_7$&$z_8$&$z_9$&$z_{10}$&$z_{11}$&$z_{12}$&$z_{13}$&$z_{14}$\\
\hline
LCDM&0.10&0.21&0.29&0.41&$0.49$&$0.58$&0.72&0.82&0.88&.99&1.53&1.97&3.07&3.57 \\
OCDM&0.11&0.18&0.31&0.41&$0.51$&$0.63$&0.69&-&0.88&1.03&1.47&2.03&3.13&3.53 \\
\hline
\end{tabular}
\end{center}
\end{table*}

A total of 14 source redshifts were selected in each cosmology to give
good statistical coverage of the redshifts of interest (see Table 2). 
Each complete pass through all the simulation volumes and for each
source redshift was performed a total of $N=10$ times and the values
for the convergence were smoothed on the different angular scales
using a top-hat filter. The computed values for the pdfs and the
higher-order moments from each of the runs in each cosmology were
averaged so that the errors on the means of $1\sigma/\sqrt{N}$ for
each statistic were determined.

\section{Results}
\label{Results}

Comparison of our analytical results against numerical simulations 
can be divided in two different categories. In addition to comparing 
the analytical results for lower order moments we also compute the whole 
pdf $\cP(\kappa_s)$ of the smoothed convergence $\kappa_s$
for a wide range of smoothing angles and source redshifts.
This allows us to assess the accuracy of the simple parameterization
described in Sect.~\ref{A simple parameterization for the generating function}
and of the various methods presented in Sect.~\ref{Computation of the pdfs}.

\subsection{Amplitude of the convergence}
\label{Amplitude of the convergence}

\begin{figure}
\protect\centerline{
\epsfysize = 1.8truein
\epsfbox[25 400 588 715]
{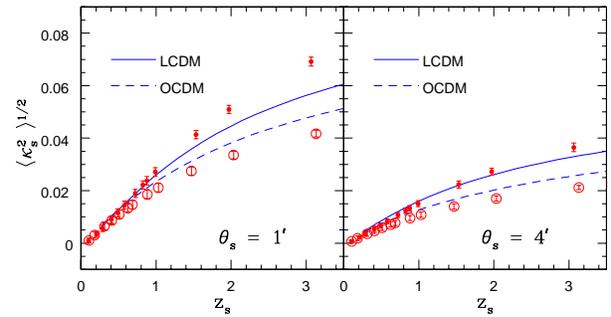}}
\caption{Variance of the smoothed convergence field $\kappa_s$, 
$\lag \kappa_s^2 \rag^{1/2}$, as a function of the source redshift $z_s$. 
The smoothing angle $\theta_s$ is fixed at $1$ arcminute in left panel and 
at $4$ arcminute in the right panel. Lines correspond to the analytical 
prediction (\ref{xikap1}) and data points represent results from numerical 
simulations (filled circles for LCDM and open circles for OCDM). 
Error bars are computed from scatter among various realizations.}
\label{Varz}
\end{figure}

\begin{figure}
\protect\centerline{
\epsfysize = 1.8truein
\epsfbox[25 400 588 715]
{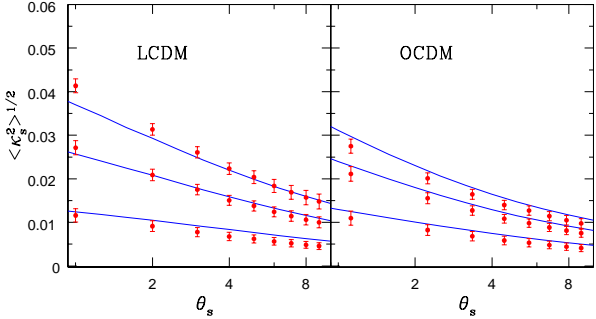}}
\caption{Variance of the smoothed convergence field $\kappa_s$, 
$\lag \kappa_s^2 \rag^{1/2}$, as a function of smoothing angle $\theta_s$ 
for three source redshifts $z_s$. These are $z_s=0.5$ (lower curve), 
$z_s=1$ (middle curve) and $z_s=1.5$ (upper curve).
Errorbars are computed from scatter among various realizations.
Different panels correspond to different cosmologies as indicated.}
\label{Vartheta}
\end{figure}

\begin{figure}
\protect\centerline{
\epsfysize = 2.0truein
\epsfbox[290 420 588 715]
{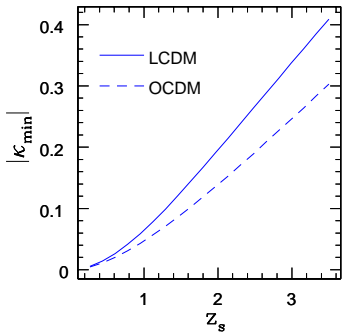}}
\caption{Minimum value $|\kappamin|$ of convergence $\kappa$ as a function 
of source redshift $z_s$. Note that $\kappamin$ is independent of smoothing
angle $\theta_s$.}
\label{Figkappamin}
\end{figure}

We first study here the amplitude of the smoothed convergence as a
function of angle and source redshift, for both LCDM and OCDM
cosmologies.  Thus, we show in Fig.~\ref{Varz} the dependence on the
source redshift $z_s$ of the {\it rms} convergence $\lag
\kappa_s^2\rag^{1/2}$ for two smoothing angles, $\theta_s=1'$ and
$\theta_s=4'$. The lines show the analytical prediction (\ref{xikap1})
while the data points are the results from numerical simulations. The
{\it rms} convergence increases with the source redshift, $z_s$,
according to the (cosmology-dependent) rate of formation of structure
and the location of massive structures for their gravitational lensing
effects, as described by Barber et al., 2000. This makes the pdf
broader. The variance is smaller for the OCDM case mainly because the
normalization, $\sigma_8$, of the power-spectrum, $P(k)$, is smaller
(see Table 1). In a similar fashion, $\lag \kappa_s^2\rag^{1/2}$
decreases for larger smoothing angles which probe larger scales, see
eq.(\ref{ks}), where the amplitude of the density fluctuations is
smaller. We obtain a reasonable agreement with the results from
numerical simulations which shows that the prescription from Peacock
\& Dodds (1996) is sufficient to reproduce the non-linear
power-spectrum over the regime probed by these angles and source
redshifts.  Note indeed that eq.(\ref{xikap1}) for the variance
$\xikap$ only involves the Born and small-angle approximations, as
well as our prescription for $P(k)$ (here taken from Peacock \& Dodds,
1996).

Next, we plot in Fig.~\ref{Vartheta} the dependence on the smoothing
angle, $\theta_s$, of the {\it rms} convergence, $\lag
\kappa_s^2\rag^{1/2}$, for the three source redshifts, $z_s=0.5, 1$
and $1.5$. The redshift, $z_s$, corresponding to the curves increases
from bottom to top, in agreement with Fig.~\ref{Varz}. Of course, as
in Fig.~\ref{Varz} we can check that the variance grows at smaller
angular scales which probe deeper within the non-linear regime of
gravitational clustering. Consistently with Fig.~\ref{Varz} we again
obtain a reasonable agreement with the results from numerical
simulations, although we seem to overestimate somewhat the variance
for the OCDM cosmology. However, this overestimate is within the
errors associated with $N-$body simulations and the Peacock \& Dodds
(1996) fit to the non-linear power-spectrum. We note that the Peacock
\& Dodds (1996) fit underestimates the more recent simulation results
of Smith et al. (2002) in the range $1< \Delta^2(k) < 50$ for the LCDM
and OCDM cosmologies, whilst it overestimates their numerical data for
higher wavenumbers. Our overestimate might be due to the fact that the
slope of the power, $\Delta^2(k)$, is somewhat steeper for OCDM (see
Smith et al., 2002, Figure 15), which means that the relative
contribution from high wavenumbers should be larger; whence the
Peacock \& Dodds (1996) fit would be expected to overestimate the
simulation results by a larger amount for the OCDM cosmology than the
LCDM.

Finally, we show in Fig.~\ref{Figkappamin} the lower-bound $\kappamin$ of
the convergence $\kappa$ as a function of the source redshift $z_s$, see
eq.(\ref{kappamin}). Of course, $|\kappamin|$ increases at higher $z_s$ with
the length of the line of sight. It is interesting to compare $|\kappamin|$
shown in Fig.~\ref{Figkappamin} with the {\it rms} convergence 
$\lag \kappa_s^2\rag^{1/2}$ shown in Fig.~\ref{Varz}. Indeed, it partly 
describes the deviation of the pdf $\cP(\kappa_s)$ from the Gaussian. For
$|\kappamin| \la \lag \kappa_s^2\rag^{1/2}$ the lower-bound $\kappamin$
has a strong influence on the shape of the pdf which has to be significantly
different from the Gaussian. On the contrary, for 
$|\kappamin| \gg \lag \kappa_s^2\rag^{1/2}$ this lower-bound is located
very far in the low convergence tail of the pdf which can therefore look
roughly similar to a Gaussian (the asymmetry is weaker). Comparing 
Fig.~\ref{Figkappamin} with Fig.~\ref{Varz} we see that the pdf will be more
asymmetric for low source redshifts $z_s$. This trend also follows from the 
fact that low redshifts also probe the late stages of gravitational 
clustering where the density field has evolved farther from the Gaussian 
initial conditions.

\subsection{The regime of gravitational clustering probed by weak-lensing}
\label{The regime of gravitational clustering probed by weak-lensing}

\begin{figure}
\protect\centerline{
\epsfysize = 2.0truein
\epsfbox[26 420 314 715]
{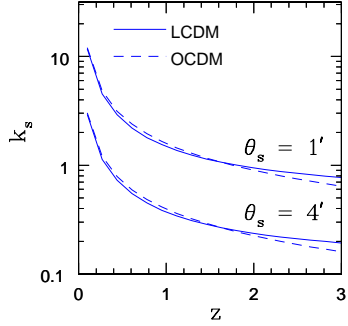}}
\caption{The typical comoving wavenumber $k_s$ (in units of $h$Mpc$^{-1}$) 
probed by the smoothed convergence
$\kappa_s$ as a function of the redshift $z$ along the line of sight, from
eq.(\ref{ks}). The upper pair of curves corresponds to the smoothing angle 
$\theta_s = 1'$ whereas the lower set of curves corresponds to the smoothing 
angle $\theta_s = 4'$. Note that the redshift $z$ is not the source redshift
$z_s$ but the redshift along the line of sight ($0<z<z_s$) of the intermediate
lensing structures.}
\label{Figks}
\end{figure}

\begin{figure}
\protect\centerline{
\epsfysize = 2.0truein
\epsfbox[305 420 588 715]
{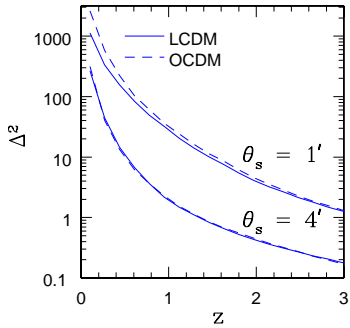}}
\caption{The power per logarithmic wavenumber interval $\Delta^2(k_s(z),z)$ 
as a function of redshift $z$ along the line of sight. The upper pair of 
curves corresponds to 
the smoothing angle $\theta_s = 1'$ while the lower set of
curves corresponds to the smoothing angle $\theta_s = 4'$.}
\label{FigDelta}
\end{figure}

\begin{figure}
\protect\centerline{
\epsfysize = 2.0truein
\epsfbox[305 420 588 715]
{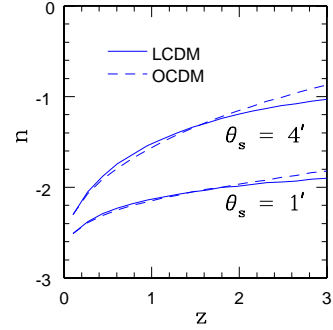}}
\caption{Spectral index $n(k_s(z))$ as a function of redshift $z$ along 
the line of sight. The upper pair of curves corresponds to the smoothing 
angle $\theta_s = 1'$ while the lower set of curves corresponds to the 
smoothing angle $\theta_s = 4'$.}
\label{Fignspec}
\end{figure}

As the smoothing angle $\theta_s$ and the source redshift $z_s$ vary the
convergence $\kappa_s$ probes different scales and different regimes of 
gravitational clustering. This could allow one to derive some information
about the physics of the gravitational dynamics in the expanding universe
from future weak-lensing surveys, in addition to the measure of the main
cosmological parameters. 

First, because of the dependence on redshift of the angular distance 
$\De(z)$, eq.(\ref{De}), the typical comoving wavenumber $k_s(z)$
probed by the smoothed convergence $\kappa_s$ varies with the redshift $z$
along the line of sight, see eq.(\ref{ks}). Thus, we display in 
Fig.~\ref{Figks} this typical comoving wavenumber $k_s$ as a function 
of $z$, for
both LCDM and OCDM cosmologies and both smoothing angles $\theta_s = 1'$
and $\theta_s = 4'$. Note that the redshift $z$ is not the source redshift
$z_s$ but the redshift along the line of sight ($0<z<z_s$) of the intermediate
lensing structures which give rise to the deflection of the light rays. As seen from
eq.(\ref{ks}), the wavenumber $k_s$ is actually proportional to $1/\theta_s$
as larger smoothing angles probe larger scales and smaller wavenumbers.
We see that for $\theta_s = 1'$ the typical comoving wavenumbers probed by the 
convergence are of order $1$ to $10 h$Mpc$^{-1}$ which corresponds to scales
$0.1$ to $1 h^{-1}$Mpc. These are the scales of present galaxies and clusters.
In particular, this means that for such angular scales weak lensing mainly
probe the intermediate regime of gravitational clustering.

This is clearly apparent in Fig.~\ref{FigDelta} which shows that the 
power per logarithmic wavenumber interval $\Delta^2(k_s(z),z)$ 
at the scales probed by weak-lensing typically runs from 
$\Delta^2(k_s(z),z) \sim 1$ up to $\Delta^2(k_s(z),z) \sim 400$. Note that
the properties of the density field show a fast evolution in this transition
regime. This entails a critical test of our simple parameterization described
in Sect.~\ref{A simple parameterization for the generating function}, which 
must be able to follow the evolution of gravitational clustering from linear
to highly non-linear scales. Moreover, we can suspect the mean-redshift
approximation (\ref{zero1}) to fail to reproduce the results from numerical
simulations with a high accuracy in this transition range. Indeed, the value
$\varphi(y;z_*)$ of the generating function at the typical redshift $z_*$ may
be a low-accuracy approximation to the mean (\ref{phispher1}) of the 
generating functions $\varphi(y;z)$ which characterize the density 
fluctuations encountered along the line of sight, as the latter show a
significant evolution from $z=0$ up to $z=z_s$. This point will appear clearly
in Sect.~\ref{The skewness of the smoothed convergence} where we discuss
the skewness of the convergence shown in Fig.~\ref{FigS3theta}.

Finally, we show in Fig.~\ref{Fignspec} the slope $n$ of the linear 
power-spectrum at the scales probed by the smoothed convergence along 
the line of sight. We see that it runs from $n \simeq -2.5$ (for small 
angles and redshifts) up to $n \simeq -1$ (for large angles and redshifts). 
Together with Fig.~\ref{FigDelta}, this shows
that gravitational weak-lensing actually probes a wide range of physical
conditions for the underlying density field. This requires the use of
flexible models which can cover any hierarchical power-spectrum and follow
the gravitational dynamics from the quasi-linear regime up to the highly 
non-linear regime, for arbitrary cosmological parameters. Fortunately, we
shall see in the following sections that the simple model presented in
Sect.~\ref{A simple parameterization for the generating function}, which we
use in this article (except for the log-normal approximation introduced in
Sect.~\ref{Lognormal approximation}), is able to recover the results
obtained from numerical simulations with a good accuracy over all ranges of
interest. On the other hand, from an observational point of view, we note that
this feature makes it somewhat more difficult to derive precise constraints
on the non-linear gravitational dynamics from weak-lensing surveys. Indeed,
it is difficult to extract one precise scale and gravitational regime from
the observed smoothed convergence $\kappa_s$.

\subsection{The skewness $\Skewkaps$ of the smoothed convergence}
\label{The skewness of the smoothed convergence}

\begin{figure}
\protect\centerline{
\epsfysize = 1.8truein
\epsfbox[25 400 588 715]
{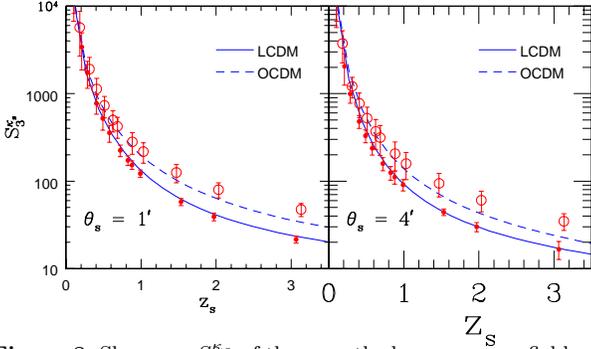}}
\caption{Skewness $\Skewkaps$ of the smoothed convergence field $\kappa_s$
as a function of the source redshift $z_s$. The left panel corresponds to the
smoothing angle $\theta_s=1'$ whereas the right panel is for $\theta_s=4'$.
The lines show our analytical prediction (\ref{S3kapspher}) from the 
spherical-cell
approximation while data points represent results from numerical simulations. 
Error bars are computed from the scatter among various realizations.}
\label{FigS3z}
\end{figure}

\begin{figure}
\protect\centerline{
\epsfysize = 1.8truein
\epsfbox[25 400 588 715]
{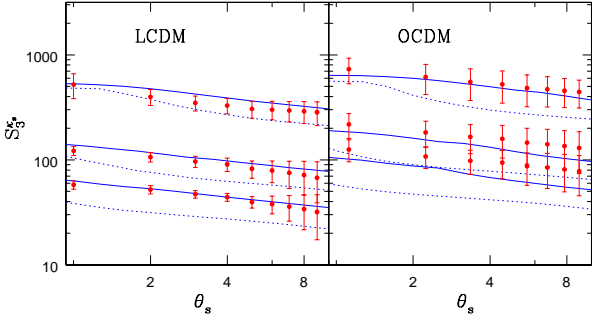}}
\caption{Skewness $\Skewkaps$ of the smoothed convergence field $\kappa_s$
as a function of smoothing angle $\theta_s$ (in arcminute) for three source 
redshifts $z_s$.
These are $z_s=0.5$ (upper curve), $z_s=1$ (middle curve) and $z_s=1.5$ 
(lower curve). Different panels correspond to different cosmologies as 
indicated. The solid lines show the spherical-cell approximation 
(\ref{S3kapspher}) whereas the dotted lines correspond to the mean-redshift 
approximation (\ref{S3kapmean}). The predictions from the stellar
model are almost identical to spherical-cell approximation and are not
reproduced here for clarity.}
\label{FigS3theta}
\end{figure}

The simplest measure of the progress of gravitational clustering from small
Gaussian initial conditions is provided by the third-order moment of the
density field $\lag \delta_R^3\rag$. As discussed in 
Sect.~\ref{Cumulant generating function for the density contrast}, it is
actually more convenient to study the skewness $S_3$ defined by the ratio
$S_3= \lag \delta_R^3\rag/\lag \delta_R^2\rag^2$, see eq.(\ref{phidel2}).
Since the convergence is linear over the density field, see eq.(\ref{kappa}),
it is natural to consider also the skewness $\Skewkaps$ of the smoothed 
convergence $\kappa_s$, defined in the same fashion. Thus, we show in 
Fig.~\ref{FigS3z} the dependence of $\Skewkaps$ on the source redshift $z_s$.
For clarity, we only plot the result from the spherical-cell approximation,
which yields from eq.(\ref{cum2}):
\beq
\mbox{spherical-cell} : \hspace{0.3cm} 
\Skewkaphs = \frac{\lag \kaphs^3\rag}{\lag \kaphs^2\rag^2} = 
\frac{\int \d\chi \; \wh^3 \; S_3(z) \; \Ikap^2}
{\left[ \int \d\chi \; \wh^2 \; \Ikap \right]^2}
\label{S3kapspher}
\eeq
and:
\beq
\Skewkaps = \frac{\lag \kappa_s^3\rag}{\lag \kappa_s^2\rag^2} =
\frac{\Skewkaphs}{|\kappamin|} .
\label{S3kapspherical}
\eeq
We consider both LCDM and OCDM cosmologies, and the two smoothing angles
$\theta_s = 1'$ and $\theta_s = 4'$. We can check in the figure that the
skewness $\Skewkaps$ decreases for larger source redshift $z_s$. This is
partly due to the fact that higher redshifts probe earlier stages of 
gravitational clustering where the density field is closer to Gaussian and 
its skewness $S_3$ is also smaller. However, most of this decrease is due
to the sum along the line of sight over successive mass sheets. Indeed,
the total convergence $\kappa$ is the integral over the whole line of sight
of the random contributions associated to the successive lens planes at
redshifts $0<z<z_s$, see eq.(\ref{kappa}). Then, this sum of random variables
tends to make the total signal closer to Gaussian through the central limit
theorem. This can also be seen as follows. Within the mean-redshift 
approximation (\ref{zero1}) we have for the skewness $\Skewkaphs$ of the 
normalized smoothed convergence $\kaphs$ and for $\Skewkaps$:
\beq
\mbox{mean-redshift} : \hspace{0.2cm}
\Skewkaphs = S_3(z_*), \hspace{0.2cm} \mbox{and} \hspace{0.2cm}
\Skewkaps = \frac{S_3(z_*)}{|\kappamin|} .
\label{S3kapmean}
\eeq
Thus, we see from eq.(\ref{S3kapmean}) and from the increase of $|\kappamin|$
with the source redshift $z_s$ shown in Fig.~\ref{Figkappamin}, that even
for a constant skewness $S_3(z)$ for the density field we would obtain
a decrease with the source redshift for the skewness $\Skewkaps$ of the
smoothed convergence (while the skewness $\Skewkaphs$ of the normalized 
smoothed convergence $\kaphs$ would remain constant within the mean-redshift 
approximation). Of course, as noticed above, this trend is actually reinforced
by the fact that higher redshifts also probe density fields which are closer
to Gaussian (i.e. $S_3(z)$ declines at larger $z$). From the behaviour of the
skewness $\Skewkaps$ we can already infer that the pdf $\cP(\kappa_s)$ of the
convergence will look closer to Gaussian for higher source redshifts.

We can see in Fig.~\ref{FigS3z} that our analytical prediction 
(\ref{S3kapspher}) shows a good agreement with the results from numerical
simulations. This means that our simple parameterization 
(\ref{S31})-(\ref{S32}) for the skewness of the density field provides
a reasonable estimate over the range probed by weak-lensing. Indeed, the
prediction obtained from the stellar model approximation is almost identical
to the result computed from eq.(\ref{S3kapspher}) which means that the
spherical-cell approximation is very good (at least for low-order moments
like the skewness) and that the comparison with numerical simulations
in Fig.~\ref{FigS3z} mainly tests our simple parameterization 
(\ref{S31})-(\ref{S32}) for the skewness of the density field.
Note the rather small dependence on $\Ol$ of the skewness $\Skewkaps$.
On the other hand, the skewness is well-known to show a strong dependence
on the cosmological parameter $\Om$, which explicitly appears in $\kappamin$,
see eq.(\ref{kappamin}). This provides a good tool to measure $\Om$ from
weak-lensing surveys, see Bernardeau et al. (1997).
 
Next, we plot in Fig.~\ref{FigS3theta} the dependence on the smoothing
angle $\theta_s$ of the skewness $\Skewkaps$ for the three source
redshifts $z_s=0.5,1$ and $1.5$. In agreement with Fig.~\ref{FigS3z}
the redshift $z_s$ corresponding to the curves increases from top down
to bottom. The solid lines are the spherical-cell approximation
(\ref{S3kapspher}) while the dotted lines are the mean-redshift
approximation (\ref{S3kapmean}). We can see in Fig.~\ref{FigS3theta}
that the skewness $\Skewkaps$ decreases for large smoothing angles
$\theta_s$. This is due to the fact that large angles probe large
scales which are closer to the linear regime, whence their skewness
$S_3$ is smaller. This trend is also reinforced by the fact that for
smaller wavenumbers $k$ the local slope $n$ of the linear
power-spectrum increases (for CDM-like power-spectra like those we
study here) which leads to a smaller skewness $S_3$ in both the linear
and highly non-linear regimes, see eq.(\ref{S31}). We can check that
our prediction (\ref{S3kapspher}) agrees reasonably well with the
numerical simulations which shows that our simple parameterization
(\ref{S31})-(\ref{S32}) works fairly well. The results obtained from
the stellar model approximation are almost identical to the solid
lines which means that the spherical-cell approximation is quite
good. In fact, the inaccuracy is dominated by far by the
parameterization (\ref{S31})-(\ref{S32}) rather than by the
spherical-cell approximation.

On the other hand, we note that the 
mean-redshift approximation (\ref{S3kapmean}) shows some 
discrepancies with the simulation results. In particular, it yields a
characteristic step-like profile with a sharp decrease for larger angles.
This step corresponds to the transition to the highly non-linear regime where
the skewness suddenly shows a sharp variation, as it evolves from 
$S_3^{\rm QL}$ up to $S_3^{\rm NL}$. The plateau at small angles corresponds
to the highly non-linear regime where the skewness saturates, within our
approximation (\ref{S31}), and the slope $n$ of the linear power-spectrum 
shows a very weak dependence on scale. Therefore, we clearly see in 
Fig.~\ref{FigS3theta} that for smoothing angles $\theta_s \sim 4'$ and 
source redshifts $z_s \sim 1$ the convergence $\kappa_s$ actually probes
the intermediate regime of gravitational clustering, see also 
Fig.~\ref{FigDelta}. This sharp feature does not show up in the prediction
(\ref{S3kapspher}) given by the spherical-cell approximation because the latter
involves an integration over redshift along the line of sight. This makes the
prediction for $\Skewkaps$ smoother and the decline at larger angles is
shallower since it takes into account the highly non-linear scales which are
still probed at low redshifts $z<z_*$, see also the rise at low $z$ of the
typical wavenumber $k_s$ shown in Fig.~\ref{Figks}. This shows that the
mean-redshift approximation presented in 
Sect.~\ref{Zeroth-order approximation} is not sufficiently accurate to obtain
a good estimate of the skewness $\Skewkaps$. Of course, at very small angles,
where the skewness $S_3$ of the density field at the typical wavenumber
$k_s(z)$ only shows a weak variation along the line of sight, as discussed
above, the mean-redshift approximation becomes quite good and it recovers
the result of eq.(\ref{S3kapspher}). 

We may note that at small angles the results of numerical simulations
seem to keep rising rather than to show a plateau as for the
analytical predictions. This suggests that our prescription
(\ref{S31}) is only approximate and that the skewness keeps slowly
increasing within the highly non-linear regime. This corresponds to a
small deviation from the stable-clustering {\it Ansatz}. Note that the
latter is actually a lower bound in the sense that the skewness $S_3$
must either remain constant or increase in the highly non-linear
regime as shown in Valageas (1999) (this only follows from the fact
that the matter density is positive). An alternative to our simple
parameterization (\ref{S31}) would be to use a halo model to evaluate
$S_3$ within the non-linear regime. However, we shall not investigate
such a model in this article since the simple parameterization
(\ref{S31}) already provides a reasonable match to the simulation
results over the range of interest, as seen in Fig.~\ref{FigS3theta},
and, in addition, our prescription seems to yield a better match to
numerical simulations than the halo-model used in Takada \& Jain (2002).

Finally, let us note that higher order moments such as the skewness take 
contributions from the high-$\kappa_s$ tail of the pdf. Hence they are more 
affected by the finite size of the cutoff than the variance.
However we believe that for the angular scales considered in our calculations
we are not limited by the size of the catalogue. We have used several
realizations to probe the underlying mass distribution which provides
a good handle on the errors introduced sample variance.

\subsection{The pdf $\cP(\kappa_s)$ of the smoothed convergence}
\label{The pdf}

\begin{figure}
\protect\centerline{
\epsfysize = 3.truein
\epsfbox[25 140 588 715]
{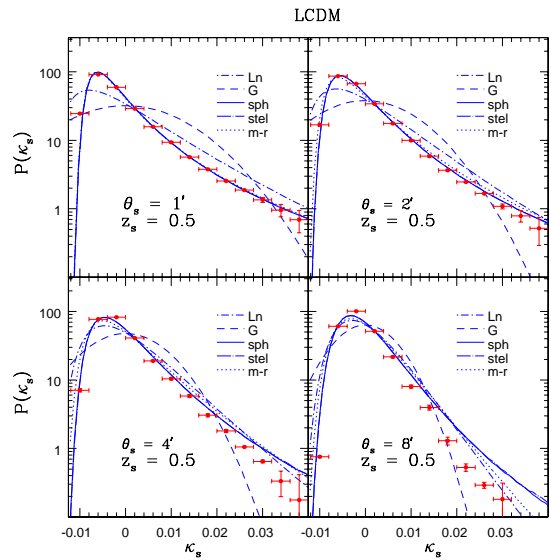}}
\caption{Probability distribution function (pdf) of the smoothed 
convergence $\kappa_s$, $\cP(\kappa_s)$, for an LCDM cosmology. 
Various panels correspond to various smoothing angles. The sources are fixed 
at a redshift $z_s =0.5$. Lines correspond to the analytical predictions
as labeled in the Figure: spherical-cell approximation (\ref{phispher1}) 
(solid line), the mean-redshift approximation (\ref{zero1}) (dotted line), 
the log-normal approximation (\ref{ln1}) (dot-dash line), the stellar-model 
approximation (\ref{phistel1}) (dot-long dash line) and the Gaussian 
(dashed line). The predictions obtained from the spherical-cell approximation
and the stellar-model are almost indistinguishable.
Data points show the results from numerical simulations.}
\label{pdfLz05}
\end{figure}

\begin{figure}
\protect\centerline{
\epsfysize = 3.truein
\epsfbox[25 140 588 715]
{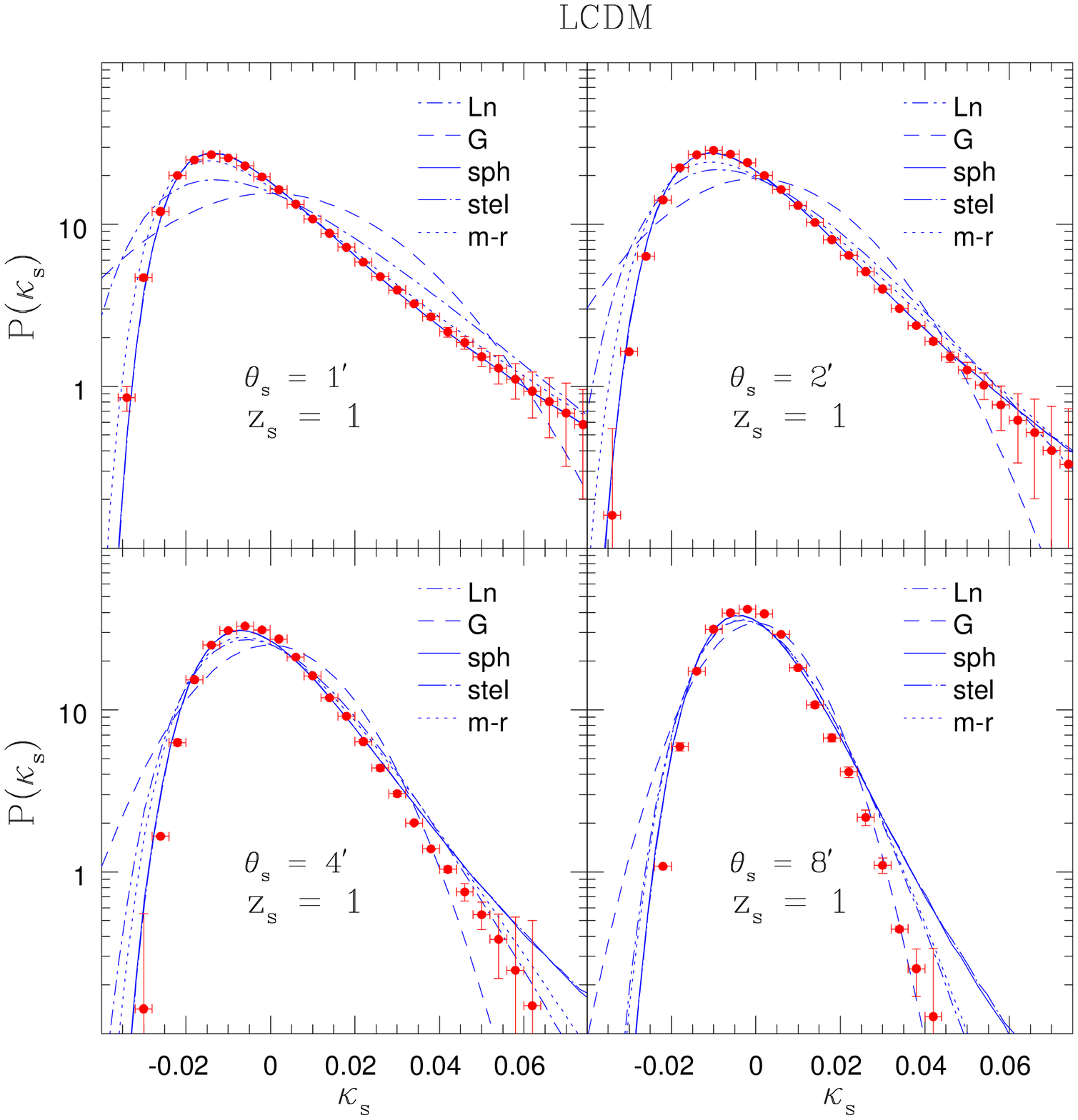}}
\caption{As for the previous Figure, but for source redshift $z_s =1$.}
\label{pdfLz1}
\end{figure}

\begin{figure}
\protect\centerline{
\epsfysize = 3.truein
\epsfbox[25 140 588 715]
{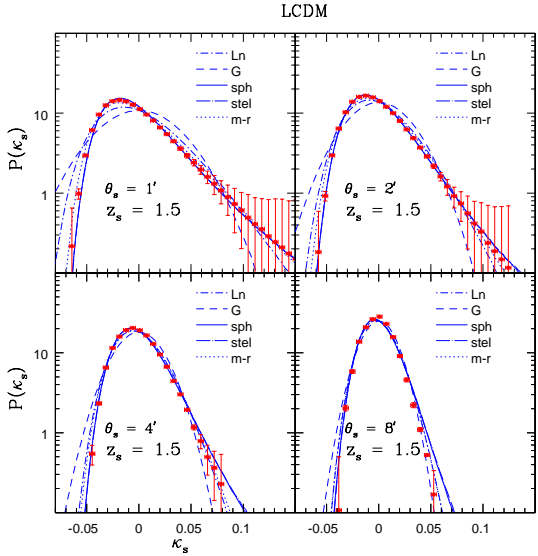}}
\caption{As for the previous Figure, but for source redshift $z_s=1.5$.
Notice that although the variance of $\kappa_s$ increases with source redshift
the distribution tends to become more Gaussian (see text for discussion).}
\label{pdfLz15}
\end{figure}

\begin{figure}
\protect\centerline{
\epsfysize = 3.truein
\epsfbox[25 140 588 715]
{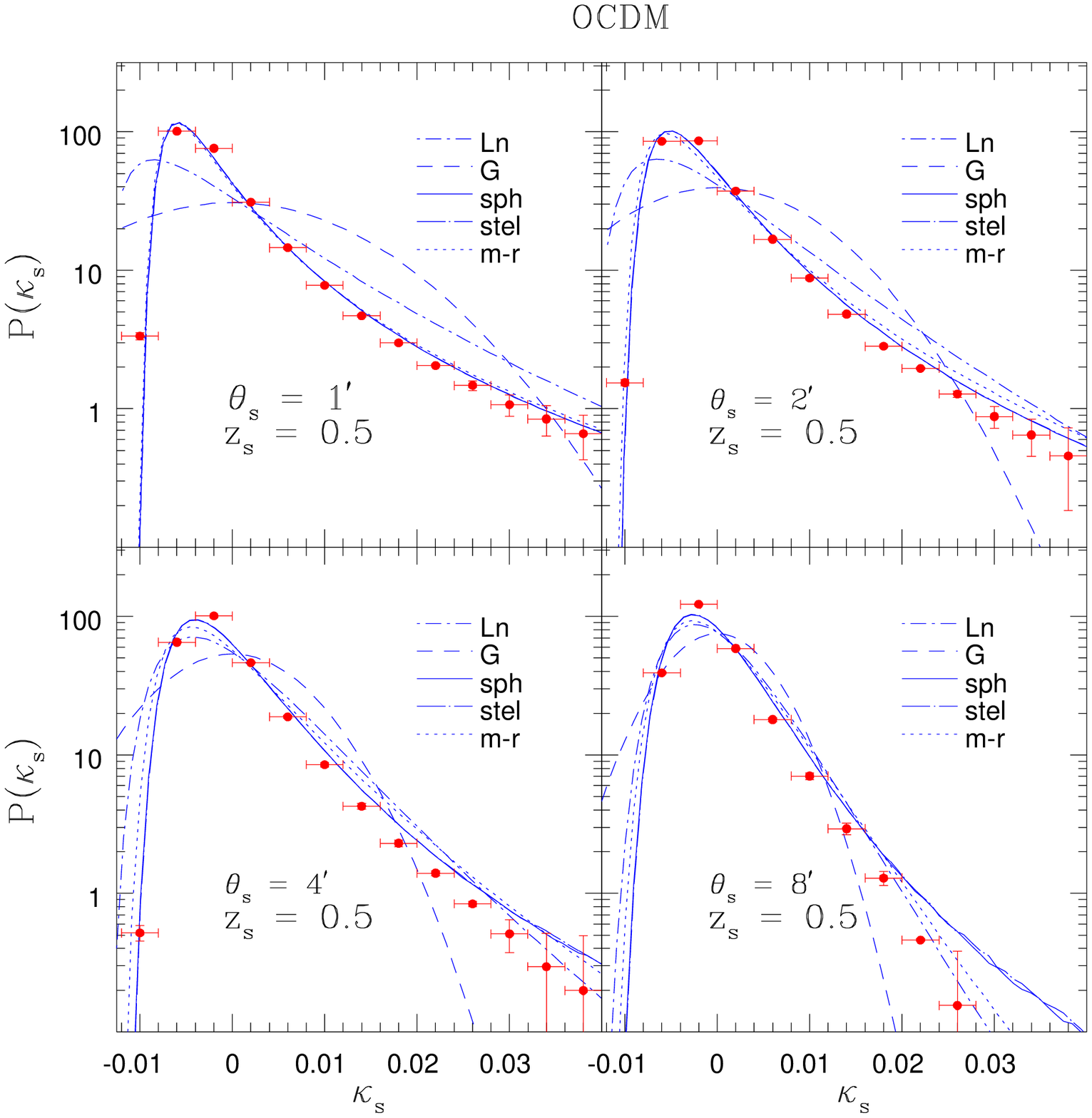}}
\caption{Probability distribution function (pdf) of the smoothed 
convergence $\kappa_s$, $\cP(\kappa_s)$, for an OCDM cosmology. 
Various panels correspond to various smoothing angles. The sources are fixed 
at a redshift $z_s =0.5$. Lines correspond to the analytical predictions
as labeled in the Figure: spherical-cell approximation (\ref{phispher1}) 
(solid line), the mean-redshift approximation (\ref{zero1}) (dotted line), 
the log-normal approximation (\ref{ln1}) (dot-dash line), the stellar-model 
approximation (\ref{phistel1}) (dot-long dash line) and the Gaussian 
(dashed line). The predictions obtained from the spherical-cell approximation
and the stellar-model are almost undistinguishable.
Data points show the results from numerical simulations.}
\label{pdfOz05}
\end{figure}

\begin{figure}
\protect\centerline{
\epsfysize = 3.truein
\epsfbox[25 140 588 715]
{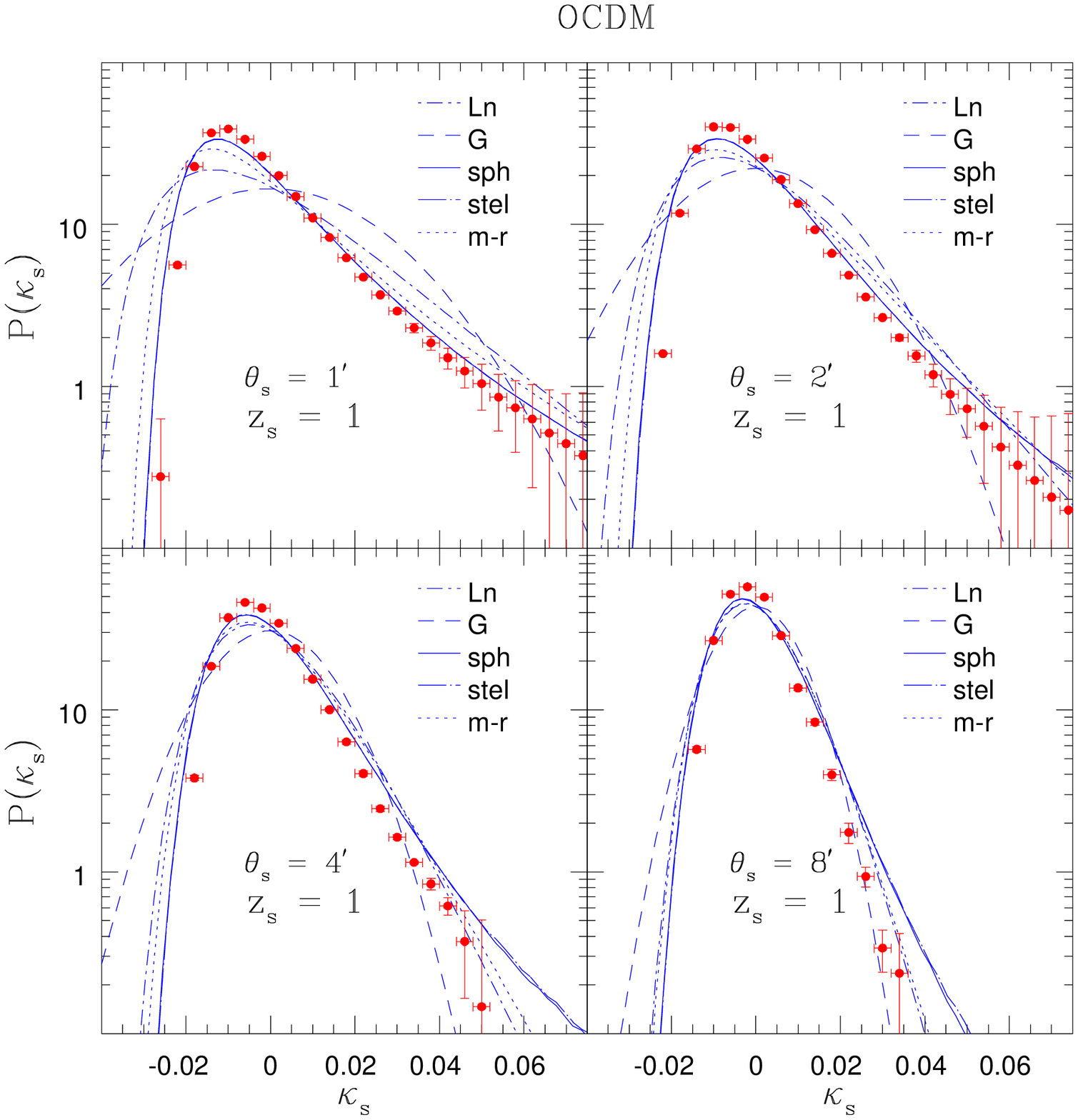}}
\caption{As for the previous Figure, but for source redshift $z_s =1$.}
\label{pdfOz1}
\end{figure}

\begin{figure}
\protect\centerline{
\epsfysize = 3.truein
\epsfbox[25 140 588 715]
{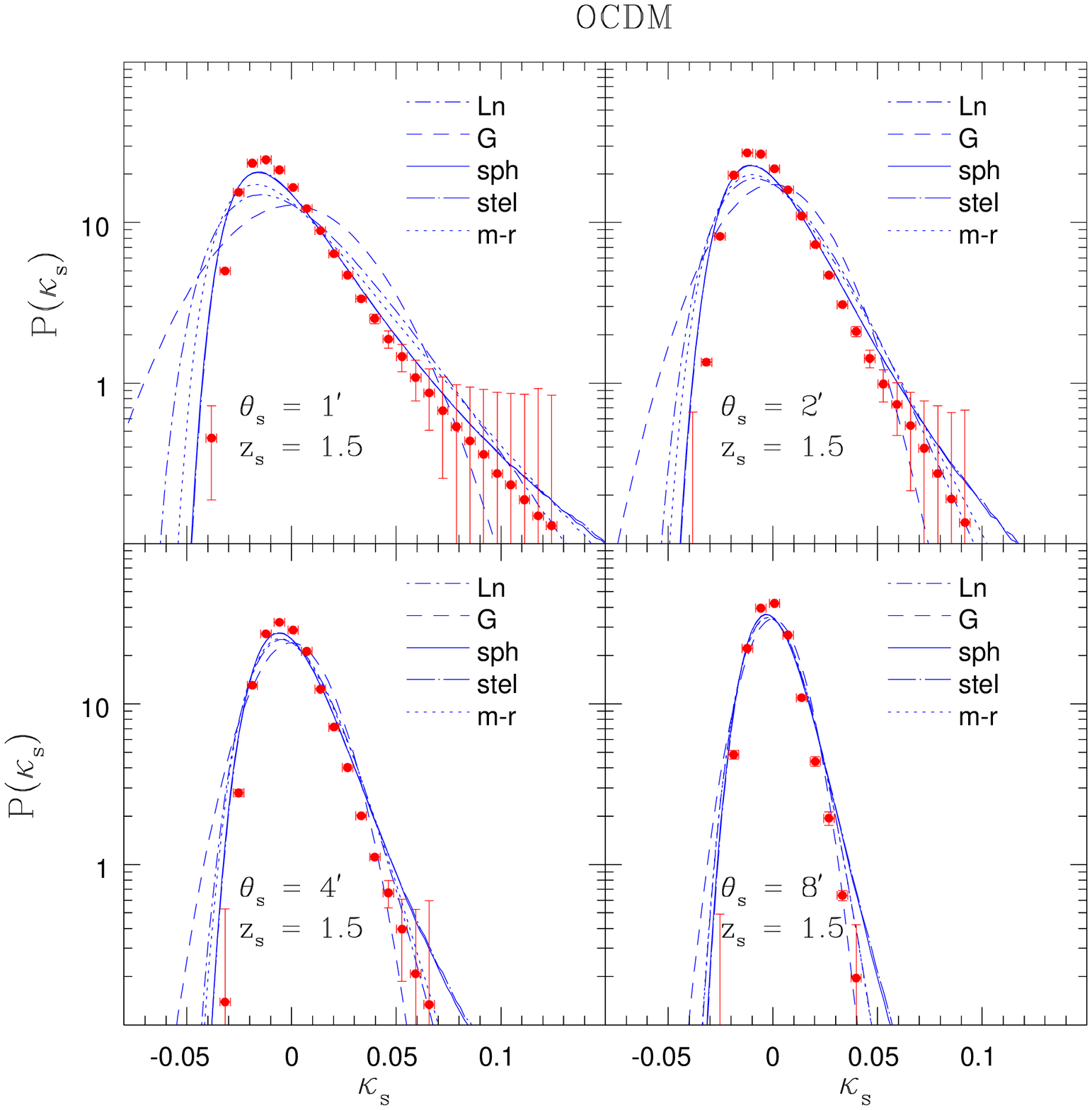}}
\caption{As for the previous Figure, but for source redshift $z_s =1.5$.}
\label{pdfOz15}
\end{figure}

Finally, we investigate here the full probability distribution function of
the smoothed convergence $\kappa_s$. Our analytical predictions mainly
depend on our prescription for the non-linear power-spectrum, taken here
from Peacock \& Dodds (1996), which was specifically tested in 
Sect.~\ref{Amplitude of the convergence}, on our model (\ref{S31})-(\ref{S32})
for the skewness, which was specifically tested in 
Sect.~\ref{The skewness of the smoothed convergence}, and on our simple
parameterization (\ref{phiMF})-(\ref{zetadef}) for higher-order cumulants.
Therefore, the shape of the pdf $\cP(\kappa_s)$ does not involve new parameters
in addition to those already tested in the previous sections against numerical
simulations through the variance and the skewness of the convergence.
Hence the comparison with simulations of the shape of the pdf itself mainly 
tests the simple form (\ref{zetadef}) we used for the function $\zeta(\tau)$
(the parameter $\kappa$ being already tested through the skewness).

We compare in Figs.~\ref{pdfLz05}-~\ref{pdfOz15} our analytical predictions
against results from numerical simulations for both LCDM and OCDM 
cosmologies. We consider four smoothing angles, $\theta_s = 1', 2', 4', 8'$
and three source redshifts $z_s = 0.5, 1,1.5$. First, in agreement with
Sect.~\ref{Amplitude of the convergence} we can check that the pdf gets
broader at higher $z_s$ as the variance increases. Second, the pdf becomes
closer to a Gaussian at higher $z_s$, in agreement with the discussion in
Sect.~\ref{Amplitude of the convergence} and with the decrease at larger
$z_s$ of the skewness seen in 
Sect.~\ref{The skewness of the smoothed convergence}.

The various line styles correspond to the different analytical methods.
Thus, we plot the spherical-cell approximation (\ref{phispher1}) (solid line),
the mean-redshift approximation (\ref{zero1}) (dotted line), the log-normal
approximation (\ref{ln1}) (dot-dash line), the stellar-model approximation 
(\ref{phistel1}) (dot-long dash line) and the Gaussian (dashed line).
We can see that the Gaussian cannot reproduce the pdf since over these
angular scales and source redshifts weak-lensing mainly probes the non-linear
regime so that the pdf is already strongly asymmetric. In particular, it shows
a sharp cutoff at low convergences $\kappa_s \ga \kappamin$ and an extended
tail at large positive convergences, which follows the shape of the pdf
$\cP(\delta_R)$ of the density contrast itself. Next, we note that although
the log-normal pdf is able to exhibit a large asymmetry and provides a 
significant improvement over the Gaussian, it usually fails to reproduce with
a reasonable accuracy the results from numerical simulations. This discrepancy
is especially clear at low angles, where the 
variance and the skewness of the density field at scales probed by 
weak-lensing are large. The log-normal approximation fares better at large
angles. This is due to two effects. First, the variance decreases and the pdf
gets closer to the Gaussian so that any sensible approximation (like the
log-normal) which goes to the Gaussian for small variance would improve in
this limit. Second, larger angles probe larger scales where the slope $n$ of
the linear power-spectrum increases and the skewness decreases and becomes
closer to $3$. Then, as noticed in Sect.~\ref{Lognormal approximation} the
log-normal approximation is actually very good in the quasi-linear regime
for $n \simeq -1$ and $S_3 \simeq 3$ (it also coincides with our
parameterization (\ref{phiMF})-(\ref{zetadef}) in this limit). This explains
why the log-normal pdf works better at large angles $\theta_s \sim 8'$.

On the other hand, we can see that our analytical predictions based on the
simple parameterization described in 
Sect.~\ref{A simple parameterization for the generating function}
show a reasonably good agreement with numerical simulations over all angular
scales and source redshifts. The improvement over the log-normal pdf is not
surprising since the model (\ref{zetadef})-(\ref{S32}) allows us to follow
the dependence on time and scale of the skewness $S_3$ of the density field for
any slope $n$ of the linear power-spectrum. This was already specifically
tested in Sect.~\ref{The skewness of the smoothed convergence}. Then,
Figs.~\ref{pdfLz05}-~\ref{pdfOz15} show that the simple form (\ref{zetadef})
for the function $\zeta(\tau)$ which implicitly determines the higher-order
moments of the density field provides a reasonable prescription. In fact,
the agreement with numerical simulations is surprisingly good in view of
the simplicity of the model (\ref{zetadef})-(\ref{S32}). At large angles
our prescription seems to overestimate the large-$\kappa_s$ tail of the pdf.
This might be cured by using for the function $\zeta(\tau)$ the exact result
derived for the quasi-linear regime (Bernardeau, 1994, Valageas, 2002), or
our simple interpolation (\ref{S32}) may overestimate the skewness in the
transition regime, although this does not seem to be the case for the OCDM
cosmology, see Fig.~\ref{FigS3theta}. However, since the high-$\kappa_s$ 
tail of the pdf for large angles $\theta \ga 8'$ may not be of great practical
interest we shall not try in this article to improve over the model 
(\ref{zetadef})-(\ref{S32}) which shows the advantage of a great simplicity.

From Figs.~\ref{pdfLz05}-~\ref{pdfOz15}, we note that all approximations 
based on the simple 
parameterization (\ref{zetadef})-(\ref{S32}) yield very close predictions.
They only show some small differences in the far tail of the pdf, at large
positive $\kappa_s$, which is very sensitive to the details of the model.
In particular, the agreement between the spherical-cell prediction 
(\ref{phispher1}) (solid line) and the stellar-model prediction
(\ref{phistel1}) (dot-long dash line) shows that the spherical-cell 
approximation (\ref{spher1}) is very accurate and it is sufficient to derive
the pdf of the smoothed convergence $\cP(\kappa_s)$. Therefore, it is not
necessary to know the detailed behaviour of the many-body correlations 
$\xi_p$: the knowledge of the cumulants $\lag\delta_R^p\rag_c$ over spherical
cells is largely sufficient to predict with a very high accuracy the pdf 
$\cP(\kappa_s)$ of the smoothed convergence $\kappa_s$. This is an important
point since it shows that the measure of the pdf $\cP(\kappa_s)$ would provide
a very robust estimate of the pdf $\cP(\delta_R)$, with no degeneracy with
the detailed angular dependence of the many-body correlations (e.g., whether
they are described by a minimal tree-model or a stellar-model).

Eventually, let us note that the numerical computation of the pdf from 
simulation maps suffers from various systematics. For a finite size catalogue, 
clearly the high-$\kappa_s$ tail cannot continue to infinity. In general 
the high-$\kappa_s$ tail shows large fluctuations due to the presence 
(or the absence) of rare overdense (underdense) objects before showing an 
abrupt cutoff. Such effects have been studied in great detail for galaxy 
catalogues (see e.g. Bernardeau et al. 2002) 
as well as for weak lensing surveys (Munshi \& Coles 2003). A comparison of 
our numerical results with larger simulation volumes will help us to quantify 
such systematic deviations.

\section{Discussion}
\label{Discussion}


Ongoing weak lensing surveys with wide field CCD imaging are being
used to produce shear maps for areas of order 10 square degrees. Soon,
larger patches, $10^{\circ} \times 10^{\circ}$, will also be feasible,
e.g., from various surveys including the MEGACAM camera on the
Canada-France-Hawaii Telescope and the VLT-Survey Telescope. Such
surveys will provide a very interesting insight into the dynamics of
the universe and the clustering in the mass distribution on small
angular scales, where the density distribution is highly non-linear.
They will provide opportunities to test our knowledge of gravitational
clustering at small length scales where no rigorous analytical results
are presently available. Indeed, unlike galaxy surveys, weak lensing
surveys provide an unbiased picture of matter clustering, as they
directly probe the gravitational potential. Results from such future
surveys will be especially useful where they provide additional data
in the form of photometric redshift information based on the
multi-colour data. In these cases the additional radial information
may assist in building up a three-dimensional picture of the dark
matter structure.


Statistics of the convergence, $\kappa_s$, can not only be studied
from the by-product of shear maps generated from weak lensing surveys,
but can also be studied from the magnification effects of clustered
matter which produce variations in the image sizes and number density
of galaxies across the sky (see, e.g., Jain, 2002). Although there are
considerable difficulties in practice in using these effects, as
described at length by Bartelmann \& Schneider (2001), use of such
results from wide field surveys can map the large-scale structure and
help us to quantify its statistics. Recent order-of-magnitude analysis
of the signal-to-noise ratio as a function of angular scale and source
redshift have suggested that well-designed forthcoming surveys will
have high signal-to-noise ratios on scales of about 0.1 arcminute to
several degrees. This will help us to probe the clustering of matter
on spatial scales of about 50 kpc to 100 Mpc.


Recent studies of weak lensing have focussed mainly on recovering the
mass power-spectrum, either from weak lensing data alone or from joint
analyses of CMB and weak lensing surveys (see e.g., Contaldi et al.,
2003, for recent estimates). Although such joint analyses can pinpoint
the cosmological parameters very effectively, the non-Gaussianities we
study here form a complementary approach and are able to break the
degeneracies in estimating the cosmological parameters from weak
lensing surveys alone.


Most previous analytical studies of weak lensing statistics can be divided 
in two categories. A majority of previous works has used a perturbative 
analysis (Bernardeau et al. 1997) which is only applicable in the 
quasi-linear regime and hence will require
a large smoothing angle. Although interesting however, such studies 
will have limited use for currently ongoing surveys as survey areas 
of order 10 square degrees are necessary to validate such smoothing angles
while still keeping effects of finite size of the survey area low on
various statistical quantities. However, given that existing CCD cameras 
typically have
diameters of $0.25^{\circ}-0.5^{\circ}$, the current weak lensing surveys
are providing us statistical information on 
small smoothing angles, of order $10'$ and less.


Therefore, some earlier studies have also investigated the non-linear regime
using a hierarchical {\em Ansatz} to compute weak
lensing statistics on smaller angular scales 
(e.g. Valageas 2000a,b, Munshi \& Jain 2000, Munshi 2000). It was shown that 
the hierarchical
ansatz when combined with extensions of perturbative calculations such as
Hyper-extended perturbation theory can provide an accurate picture of
weak lensing statistics on highly nonlinear scales. In this
paper we extend these predictions to check their validity by comparing
analytical predictions with results from numerical simulations over a large 
range of angular scales and redshifts.
Thus, we have presented two model-independent approximations which allow one
to obtain the pdf of the convergence (and all its moments). The first method,
the ``spherical-cell approximation'', simply assumes that the relative 
amplitude of the many-body density correlations smoothed at a given scale 
shows a weak dependence on the shape of the real-space filter, while the 
second one, the ``mean-redshift approximation'', makes the further 
approximation that the properties of the density field along the 
line-of-sight can be estimated by considering a typical intermediate 
redshift. Next, we have evaluated the accuracy of these approximations 
by considering a specific ``stellar model'' for the many-body density 
correlations where exact calculations can be performed. We find that in this
case the spherical-cell approximation is actually very good and can barely be 
distinguished from the exact result. This shows that the smoothed convergence
$\kappa_s$ is a very robust tool to measure the statistics of the density 
fluctuations at a given scale and it does not depend on the detailed angular
behaviour of the many-body density correlations (it only involves their
overall amplitude at this scale). On the other hand, we find that the 
``mean-redshift approximation'' provides a reasonable prediction, in agreement
with previous works (Valageas 2000b), but it does not give a very accurate
prediction for the low-order cumulants like the skewness at intermediate
angular scales, as the properties of the density field show significant
variations along the line-of-sight as it evolves from the linear to the 
highly non-linear
regime. A simple model which has been proposed to describe the density field
from linear to non-linear scales is the log-normal approximation. We have
applied this model to the derivation of the pdf of the convergence based on
the mean-redshift approximation. Then, we find that while the log-normal 
approximation may be a good description for smoothing angles $\theta_s > 4'$ 
it is not sufficiently accurate for smaller scales. This is related to the
facts that such scales are deeper in the non-linear regime and the local 
slope $n$ of the linear power-spectrum $P(k)$ is significantly different
from $-1$, so that the log-normal approximation underestimates the skewness.
Finally, we have applied the three general methods described above (i.e.,
the spherical-cell, mean-redshift and stellar-model approximations) to
a very simple model for the evolution of the density fluctuations from the 
linear to the highly non-linear regime. We find a good agreement with the
results from  numerical simulations for a wide range of source redshifts 
as well as smoothing angles, for both LCDM and OCDM cosmologies.


An alternative approach to model non-linearities in gravitational
clustering which has emerged in recent years is the ``Halo model''
(see e.g.  Cooray A. \& Sheth R for a recent review). In this approach
the dark matter is modeled as belonging to halos with a mass function
given by the Press-Schechter formalism and a spatial distribution
modeled as in Mo \& White (1996) or its variants.  When supplemented
by a radial profile for the dark halos such a prescription can be used
to compute many statistical quantities associated with the density
field. These models have also been applied to model statistics of the
convergence field. Our approach is complimentary to such methods.  It
is better-suited to the study of weak-gravitational lensing effects
which directly probe the density field (and its many-body
correlations) and do not involve the decomposition of the universe
over various classes of objects and virialized halos. In this sense,
our approach is more natural and the measures obtained from
weak-lensing surveys (e.g., the low-order moments of the convergence
or the shear) are directly related to the density correlations which
are the basic ingredients of our methods. On the other hand, halo
models can provide an interesting connection between weak-lensing
effects and the properties of virialized halos (so that one might hope
to derive for instance some constraints on the profiles of these
halos). However, in our opinion, it is not clear, until fully tested,
that they can provide a full description of weak-lensing effects which
also probe low-density regions located outside of virialized objects.


In our studies, both analytical and numerical, we find that lower
order moments are rather sensitive to the source redshift. In this
article we only considered the case where all sources are located at
the same redshift $z_s$ but it is straightforward to apply our
analytical methods to more realistic broad distributions of source
redshifts. We plan to investigate this point in future studies when
relevant numerical simulations are available. In agreement with
previous results we found that while the variance of convergence field
(independently of smoothing angle) increases with the depth of the
survey, the distribution of $\kappa_s$ tends to become more Gaussian,
thereby reducing the values of the $S_p^{\kappa_s}$ parameters.
Nevertheless, the ordering of the $S_p^{\kappa_s}$ parameters in two
different cosmologies does not change with source redshifts.  For any
given source redshift the $S_3^{\kappa_s}$ parameter is higher in the
OCDM scenario as compared to the LCDM cosmology. The $S_p^{\kappa_s}$
parameters show a rapid increase at small redshifts whereas at large
redshifts they vary at a much slower rate. Of course, perturbative
analysis of the dependence of the $S_p^{\kappa_s}$ parameters on
source redshift gives a qualitatively similar dependence. 


Although we have not taken into account the effect of the finite size of
the catalogue, it is unlikely to have any effect on the length scales
we have probed in our studies. However, the finite size is likely to
play an important role in the determination of the cosmological
parameters. It is therefore of interest to incorporate such effects in
future simulations.  The effects on lower order moments have been
studied analytically by Munshi \& Coles (2003) and we note that the
higher-order, $S_p$, parameters enter in the expressions for the
computation of errors for the lower-order moments associated with the
convergence.

Noise due to the intrinsic ellipticities of lensed galaxies will also
need to be modeled. Our analytical results provide a strong foundation
for such studies and we plan to include such features in our future
work.  The intrinsic ellipticity distribution of galaxies can be
modeled by assuming a Gaussian distribution, which depends only on a
knowledge of the variance in the distribution.  To compute the noisy
pdf it is therefore necessary to convolve the theoretical convergence
pdf with the noise pdf to produce a resultant containing information
on both the underlying mass distribution and the noise. Initial
studies in this direction have been attempted by Van Waerbeke (2000)
and Jain \& Van Waerbeke (1999).


All analytical calculations regarding weak lensing statistics rely on
the so called Born approximation (Schneider et al. 1998). However, its 
validity at non-linear scales is difficult to establish by analytical 
calculations alone.
While a weakly clustered dark matter distribution is expected to produce
small deflections of photon trajectories, it is not clear whether the
small-angle approximation can be safely used at small angular scales which
probe large density fluctuations. However, the agreement of our analytical
predictions with numerical simulations strongly suggests that the Born 
approximation can be used down to $\theta_s=0.4'$ at least and that 
cases of large deflections of photon trajectories definitely
have a small statistical significance even on these small angular scales.
In fact, the theoretical uncertainty is dominated by the inaccuracy of the
simple models used for the non-linear evolution of the density field.


Finally, let us note that our simulations are different from more popular 
ones based on ray-tracing techniques and are based on computations of
the 3-dimensional shear along the lines of sight.
The comparison of analytical and numerical results for convergence maps are
complementary to similar analysis of shear maps presented in Valageas, 
Barber \& Munshi (2003). The excellent match between analytical and simulation 
results in both cases increases our faith in both the analytical and 
numerical methods. 

\section*{acknowledgments}
This work has been supported by PPARC and the numerical work carried
out with facilities provided by the University of Sussex. AJB was
supported in part by the Leverhulme Trust. The original code for the
3-d shear computations was written by Hugh Couchman of McMaster
University. DM acknowledges the support from PPARC of grant
RG28936. It is a pleasure for DM to acknowledge many fruitful
discussions with members of Cambridge Leverhulme Quantitative
Cosmology Group, including Jerry Ostriker and Alexandre Refregier.

\end{document}